\documentclass[12pt]{article}
\usepackage{amsmath,amssymb}
\usepackage{bm}
\usepackage{color}
\usepackage{graphicx}
\usepackage{cite}

\begin{document}

\title{
	\begin{flushright}
		\ \\*[-80pt]
		\begin{minipage}{0.2\linewidth}
			\normalsize
			EPHOU-19-012\\
			HUPD1912 \\*[50pt]
		\end{minipage}
	\end{flushright}
	{\Large \bf
		$A_4$ lepton flavor model and modulus stabilization \\ 
		from $S_4$ modular symmetry
		\\*[20pt]}}

\author{
	Tatsuo Kobayashi $^{1}$
	,~Yusuke Shimizu $^{2}$
	,~Kenta Takagi $^{2}$
	,\\Morimitsu Tanimoto $^{3}$
	,~Takuya H. Tatsuishi  $^{1}$
	\\*[20pt]
	\centerline{
		\begin{minipage}{\linewidth}
			\begin{center}
				$^1${\it \normalsize
					Department of Physics, Hokkaido University, Sapporo 060-0810, Japan} \\*[5pt]
				$^2${\it \normalsize
					Graduate School of Science, Hiroshima University, Higashi-Hiroshima 739-8526}\\*[5pt]
				$^3${\it \normalsize
				Department of Physics, Niigata University, Niigata 950-2181}
			\end{center}
		\end{minipage}}
	\\*[50pt]}

\date{
	\centerline{\small \bf Abstract}
	\begin{minipage}{0.9\linewidth}
		\medskip
		\medskip
		\small
		We study the modulus stabilization in an $A_4$ model whose $A_4$ flavor symmetry is originated from the $S_4$ modular symmetry.
		We can stabilize the modulus so that the $A_4$ invariant superpotential leads to the realistic lepton masses and mixing angles.
		We also discuss the phenomenological aspect of the present model as a consequence of the modulus stabilization.
	\end{minipage}
}

\begin{titlepage}
	\maketitle
	\thispagestyle{empty}
\end{titlepage}
\newpage
\section{Introduction}

The origin of the flavor structure is one of important mysteries in particle physics.
The recent development of the neutrino oscillation experiments provides us helpful information to investigate the flavor physics.
The neutrino oscillation experiments have presented two large flavor mixing angles, which contrast with quark mixing angles.
The T2K and NO$\nu$A strongly indicate the CP violation in the neutrino oscillation~\cite{Abe:2018wpn,NOvA:2018gge}.
Thus, we are in the era to develop the flavor theory with facing the experimental data.

One of the interesting approaches to understand these phenomena is to impose non-Abelian discrete symmetries for flavors \cite{
	Altarelli:2010gt,Ishimori:2010au,Ishimori:2012zz,Hernandez:2012ra,
King:2013eh,King:2014nza,Tanimoto:2015nfa,King:2017guk,Petcov:2017ggy}.
In particular, the $A_4$ flavor model was examined extensively in the neutrino phenomenology because the $A_4$ is the minimal group including a triplet irreducible representation, which enables a natural explanation of the existence of three families of leptons \cite{
Ma:2001dn,Babu:2002dz,Altarelli:2005yp,Altarelli:2005yx,Shimizu:2011xg,Petcov:2018snn,Kang:2018txu}.
However, the origin of $A_4$ symmetry is unclear.

Geometrical symmetries of compact space in extra dimensional field theories and superstring theory can be origins of non-Abelian discrete flavor symmetries \footnote{
	In Refs.~\cite{Kobayashi:2006wq,Kobayashi:2004ya,Ko:2007dz,Beye:2014nxa,Abe:2009vi},
it was shown that stringy selection rules in addition to geometrical symmetries lead to certain non-Abelian flavor symmetries.}.
Torus compactification and orbifold compactification are simple compactifications.
These compactifications have the modular symmetry $SL(2,\mathbb{Z})$ as the geometrical symmetry.
The shape of the torus is described by the modulus $\tau$, and the modular group transforms the modulus non-trivially.
The modular group $SL(2,\mathbb{Z})$ has infinite order, but it includes finite subgroups such as 
$\Gamma_2 \simeq S_3$, $\Gamma_3 \simeq A_4$, $\Gamma_4 \simeq S_4$ and $\Gamma_5 \simeq A_5$ \cite{deAdelhartToorop:2011re}.
Furthermore, the modular group transforms zero-modes each other \cite{
	Lauer:1989ax,Lerche:1989cs,Ferrara:1989qb,
Kobayashi:2017dyu,Kobayashi:2018rad,Baur:2019kwi}.
Thus, the modular symmetry is a sort of flavor symmetries.
However, Yukawa couplings as well as other couplings are functions of the modulus, and those couplings also transform non-trivially under the modular symmetry.

Inspired by these aspects, recently a new type of flavor models was proposed based on the $A_4$ modular group \cite{Feruglio:2017spp}
in which the modular forms of the weight 2 have been constructed for the $A_4$ triplet.
The successful phenomenological results also have been obtained \cite{Criado:2018thu,Kobayashi:2018scp}.
The modular forms of the weight 2 have been also constructed for
$S_3$ \cite{Kobayashi:2018vbk},
$S_4$ \cite{Penedo:2018nmg},
$A_5$ \cite{Novichkov:2018nkm},
$\Delta(96)$, and $\Delta(384)$ \cite{Kobayashi:2018bff}.
The modular forms of the weight 1 and higher weights are also given for $T'$ doublet \cite{Liu:2019khw}.
New types of flavor models towards the flavor origin were studied extensively by use of these modular forms\cite{
	Criado:2018thu,Kobayashi:2018scp,Novichkov:2018ovf,
	deAnda:2018ecu,Okada:2018yrn,Kobayashi:2018wkl,Novichkov:2018yse,Ding:2019xna,Nomura:2019jxj,Novichkov:2019sqv,
	Okada:2019uoy,deMedeirosVarzielas:2019cyj,Nomura:2019yft,Kobayashi:2019rzp,Okada:2019xqk,Kobayashi:2019mna,
Ding:2019zxk,Okada:2019mjf,King:2019vhv,Nomura:2019lnr,Okada:2019lzv,Criado:2019tzk}.

In minimal model building, we do not need to introduce flavon fields to break flavor symmetries because flavor symmetries are broken when the value of $\tau$ is fixed.
We can realize lepton and quark masses and mixing angles by choosing a proper value of the modulus $\tau$ as well as other parameters of models.
It is important how to fix the value of $\tau$, i.e. the modulus stabilization.
The modulus value can be fixed as a minimum of scalar potential in supergravity theory.
The modular invariant supergravity theory was studied \cite{Ferrara:1989bc}\footnote{
	See for their applications e.g. \cite{Derendinger:1991hq,Ibanez:1992hc,Kobayashi:2016ovu}.}.
Indeed, the modulus stabilization was studied by assuming the $SL(2,\mathbb{Z})$ modular invariance for the non-perturbative superpotential in supergravity theory \cite{Ferrara:1990ei,Cvetic:1991qm} \footnote{
	See also \cite{Kobayashi:2016mzg}.}.

The purpose of this paper is to study the modulus stabilization and its phenomenological implications in $\Gamma_N$ flavor models.
We consider the modulus stabilization by using  the model in Ref.~\cite{Kobayashi:2019mna} as an illustrating model.
Non-Abelian discrete symmetries can be anomalous \cite{Araki:2008ek}.
(See also for anomalies of the modular symmetry in concrete models \cite{Kariyazono:2019ehj}.)
For example, $S_4$ can be anomalous and thus broken down to $A_4$ by anomalies. 
In the model of Ref.~\cite{Kobayashi:2019mna}, the $S_4$ modular symmetry is imposed at the tree level and assumed to be broken to $A_4$ by anomalies.
In this paper, we study an $A_4$ invariant superpotential of the modulus $\tau$ to stabilize it at a supersymmetric minimum of the supergravity scalar potential.
We discuss phenomenological aspects in our model.

This paper is organized as follows.
In section 2, we give a brief review on the modular symmetry and the $S_4$ anomaly. 
In section 3, we review on the $A_4$ flavor model in Ref.~\cite{Kobayashi:2019mna}.
In section 4, we study the modulus stabilization in the $A_4$ model.
In section 5, we study phenomenological aspects through the modulus stabilization in the $A_4$ model.
Section 6 is devoted to our conclusion.
Relevant representations of  $S_4$ and $A_4$ groups are presented in Appendix A.
We list the input data of neutrinos in Appendix B.
In Appendix C, we show a scenario to induce the modulus potential.

\section{Modular symmetry and $S_4$ anomaly}
\subsection{Modular symmetry}

We give a brief review on the modular symmetry and modular forms.
The torus compactification is the simplest compactification.
The modulus $\tau$ of the torus transforms under the modular transformation as
\begin{equation}
	\tau \longrightarrow \tau' = \gamma \tau = \frac{a\tau + b}{c \tau + d},
\end{equation}
where $a,b,c,d$ are integer with satisfying $ad-bc=1$.
This is the symmetry $PSL(2,\mathbb{Z})=SL(2,\mathbb{Z})/\mathbb{Z}_2$, which is denoted by $\Gamma$.

The modular symmetry is generated by two elements, $S$ and $T$: 
\begin{equation}
	S: \tau \longrightarrow -\frac{1}{\tau},\qquad 
	T: \tau \longrightarrow \tau +1.
\end{equation}
They satisfy the following algebraic relations,
\begin{equation}
	\label{eq:S-ST}
	S^2=(ST)^3=\mathbb{I}.
\end{equation}
Furthermore, we define the congruence subgroups of level $N$ as 
\begin{equation}
	\Gamma(N) = \left\{ \left( 
	\begin{array}{cc}  a & b \\ c & d 
	\end{array}
	\right)  \in  PSL(2,\mathbb{Z}),   \qquad  \left( 
	\begin{array}{cc}  a & b \\ c & d 
	\end{array}
	\right)   = \left( 
	\begin{array}{cc}  1 & 0 \\ 0 & 1 
	\end{array}
	\right)  \quad ({\rm mod}~N) \right\}.
\end{equation}
The quotient subgroups $\Gamma_N$ are given as $\Gamma_N\equiv \Gamma/\Gamma(N)$, and these are finite for $N=2,3,4,5$, i.e.
$\Gamma_2 \simeq S_3$, $\Gamma_3 \simeq A_4$, $\Gamma_4\simeq S_4$, $\Gamma_5 \simeq A_5$.
The algebraic relation  $T^N = \mathbb{I}$ is satisfied for $\Gamma(N)$ in addition to Eq.(\ref{eq:S-ST}).

We study the modular invariant supergravity theory.
We use the unit that $M_P=1$ where $M_P$ denotes the reduced Planck scale.
A typical K\"ahler potential of the modulus field $\tau$ is written as follows:
\begin{equation}
	\label{eq:Kahler-tau}
	K = - \ln [i(\bar\tau - \tau)].
\end{equation}
The K\"ahler potential transforms under the modular symmetry as
\begin{equation}
	-\ln[i(\bar\tau - \tau)] \longrightarrow -\ln[i(\bar\tau - \tau)] + \ln |c\tau + d|^2.
\end{equation}
Supergravity theory can be written by $G$,
\begin{equation}
	G = K + \ln |W|^2,
\end{equation}
where $W$ denotes the superpotential in supergravity theory.
We require that $G$ is invariant under the modular transformation.
The superpotential $W$ therefore transforms as
\begin{equation}
	W \longrightarrow \frac{W}{c\tau +d},
\end{equation}
under the modular transformation.
That is, the superpotential must be a holomorphic function of the modular weight $-1$.

Chiral matter fields $\phi^{(I)}$ with the modular weight $-k_I$ transform as
\begin{equation}
	(\phi^{(I)})_i(x) \longrightarrow (c\tau +d)^{-k_I} \rho(\gamma)_{ij}(\phi^{(I)})_j(x),
\end{equation}
under the modular symmetry, where $\rho(\gamma)_{ij}$ is a unitary matrix in $\Gamma_N$.
Their K\"ahler potential can be written as
\begin{equation}
	K^{\rm matter} = \frac{1}{[i(\bar\tau - \tau)]^{k_I}} |\phi^{(I)}|^2.
\end{equation}
Moreover, the modular forms of weight $k$ are the holomorphic functions of $\tau$ and transform as
\begin{equation}
	f_i(\tau) \longrightarrow (c\tau +d)^k \rho(\gamma)_{ij}f_j( \tau).
\end{equation}
The modular forms of $\Gamma(4)$ have been constructed by use of the Dedekind eta function, $\eta(\tau)$, in Ref.~\cite{Penedo:2018nmg}.
\begin{equation}
	\eta(\tau) = q^{1/24} \prod_{n =1}^\infty (1-q^n)~,
\end{equation}
where $q = e^{2 \pi i \tau}$.
The modular forms of the weight 2 are written by
\begin{align}
	\begin{aligned}
	Y_1(\tau) & = Y(1,1,\omega,\omega^2,\omega,\omega^2|\tau),    \\ 
	Y_2(\tau) & = Y(1,1,\omega^2,\omega,\omega^2,\omega|\tau),    \\
	Y_3(\tau) & = Y(1,-1,-1,-1,1,1|\tau),                         \\
	Y_4(\tau) & = Y(1,-1,-\omega^2,-\omega,\omega^2,\omega|\tau), \\
	Y_5(\tau) & = Y(1,-1,-\omega,-\omega^2,\omega,\omega^2|\tau), 
	\end{aligned}\label{eq:Y12345}
\end{align}
where $\omega = e^{2\pi i /3}$ and 
\begin{eqnarray}
	Y(a_1,a_2,a_3,a_4,a_5,a_6 \ \tau) &=& 
	a_1 \frac{\eta'(\tau +1/2)}{\eta(\tau +1/2)} +4a_2 \frac{\eta'(4\tau )}{\eta(4\tau )} \nonumber \\
	& &+\frac14 \sum_{m=0}^3a_{m+3}  \frac{\eta'((\tau +m) /4)}{\eta((\tau +m)/4 )}.
\end{eqnarray}
These five modular forms correspond to reducible representations of $\Gamma_4 \simeq S_4$, and these are decomposed into the ${\bf 2}$ and ${\bf 3}'$ representations under $S_4$,
\begin{equation}
	Y_{S_4 {\bf 2}}(\tau) =\left(
	\begin{array}{c}
		Y_1(\tau) \\
		Y_2(\tau) 
	\end{array}
	\right), \qquad Y_{S_4 {\bf 3}'}(\tau) =\left(
	\begin{array}{c}
		Y_3(\tau)  \\
		Y_4(\tau)  \\
		Y_5((\tau) 
	\end{array}
	\right).
\end{equation}
The generators, $S$ and $T$, are represented on the above modular forms,
\begin{equation}
	\rho(S)=\left(
	\begin{array}{cc}
		0        & \omega \\
		\omega^2 & 0      
	\end{array}\right), \qquad
	\rho(T)=\left(
	\begin{array}{cc}
		0 & 1 \\
		1 & 0 
	\end{array}\right),
\end{equation}
for ${\bf 2}$, and 
\begin{equation}
	\rho(S)=-\frac13\left(
	\begin{array}{ccc}
		-1        & 2\omega^2 & 2 \omega  \\
		2\omega   & 2         & -\omega^2 \\
		2\omega^2 & -\omega   & 2         
	\end{array}\right), \qquad
	\rho(T)=
	-\frac13\left(
	\begin{array}{ccc}
		-1        & 2\omega   & 2 \omega^2 \\
		2\omega   & 2\omega^2 & -1         \\
		2\omega^2 & -1        & 2\omega    
	\end{array}\right),
\end{equation}
for ${\bf 3}'$.
The modular forms of higher weights are obtained as the products of $Y_{S_4{\bf 2} }(\tau)$ and $Y_{S_4{\bf 3}' }(\tau)$.
See for other representations in Appendix A.


\subsection{Anomaly}

A discrete symmetry can be anomalous like a continuous symmetry \cite{
Krauss:1988zc,Ibanez:1991hv,Banks:1991xj,Araki:2008ek}.
Each element $g$ in a non-Abelian discrete group satisfies $g^N=1$, that is, the Abelian $Z_N$ subgroup. 
If all of Abelian discrete subgroups in a non-Abelian discrete group are anomaly-free, 
the whole non-Abelian symmetry is anomaly-free  \cite{Araki:2008ek}.
Otherwise, the non-Abelian symmetry is anomalous, and anomalous subgroup is broken.
Furthermore, each element $g$ is represented by a matrix $\rho(g)$.
If $\det \rho(g)=1$, the corresponding $Z_N$ is always anomaly-free.
On the other hand, if $\det \rho(g) \neq 1$, the corresponding $Z_N$ symmetry can be anomalous 
\cite{Araki:2008ek,Ishimori:2010au,Ishimori:2012zz}.

In Refs.~\cite{Ishimori:2010au,Ishimori:2012zz}, it is shown explicitly which subgroups can be anomalous in non-Abelian discrete symmetries.
The $S_4$ group is isomorphic to $(Z_2 \times Z_2) \rtimes S_3$.
The $Z_2$ symmetry of $S_3$ can be anomalous in $S_4$.
In general, the ${\bf 2}$ and ${\bf 3}$ representations as well as ${\bf 1}'$ have $\det \rho(g) =-1$ 
while the ${\bf 1}$ and ${\bf 3}'$ representations have $\det \rho(g)=1$.
Indeed, $\rho(S)$ and $\rho(T)$ for ${\bf 2}$ as well as ${\bf 3}$ and ${\bf 1}'$ have $\det (\rho(S))=\det(\rho(T))=-1$.

If the above $Z_2$ symmetry in $S_4$ is anomalous, $S_4$ is broken to $A_4$ by anomalies.
In this case, $S$ and $T$ themselves are anomalous, but $\tilde S = T^2$ and $\tilde T = ST$ are anomaly-free.
These anomaly-free elements satisfy 
\begin{equation}
	(\tilde S)^2 = (\tilde S \tilde T)^3 = (\tilde T)^3 = \mathbb{I},
\end{equation}
if we impose $T^4 = \mathbb{I}$, that is, the $A_4$ algebra is realized.
The modular forms of weight 2 for $S_4$ correspond to $A_4$ representations as follows:
\begin{equation}
	Y_{S_4{\bf 2}}(\tau)\rightarrow
	(\ Y_{A_4{\bf 1'} }(\tau), \ Y_{A_4{\bf 1''}}(\tau)\ ) \ , \qquad 
	Y_{S_4{\bf 3}' }(\tau) \rightarrow Y_{A_4{\bf 3} }(\tau) \ .
\end{equation}
We have 
\begin{equation}
	Y_{A_4{\bf 1}'}(\tau)=Y_2(\tau) , \qquad Y_{A_4{\bf 1}''}(\tau)=Y_{1}(\tau), 
	\qquad
	Y_{A_4{\bf 3} }(\tau)= 
	\begin{pmatrix}
		Y_3 (\tau) \\
		Y_4 (\tau) \\
		Y_5(\tau)  
	\end{pmatrix}.
	\label{modularA4}
\end{equation}
Note that these are not modular forms of $\Gamma(3)$
because $\tilde S =T^2$ and $\tilde T= ST$ do not generate $SL(2,\mathbb{Z})$.
We can also write $S_4$ singlet modular forms of weights 4 and 6 
\begin{equation}
	Y^{(4)}(\tau) = Y_1(\tau) Y_2(\tau), \qquad Y^{(6)}(\tau) = (Y_1(\tau))^3 + (Y_2(\tau))^3.
	\label{singlets}
\end{equation}
Both are trivial singlets ${\bf 1}$ also under $A_4$. 
These are useful for our study.


\section{$A_4$ lepton model from  $S_4$ modular symmetry}

We briefly review on the $A_4$ lepton flavor model in Ref.~\cite{Kobayashi:2019mna}.
Our $A_4$ flavor symmetry is originated from the $S_4$ modular symmetry by assuming that the $S_4$ symmetry is broken to $A_4$ by anomalies as mentioned in the previous section.

The model in this paper is described in the supergravity basis where the superpotential has the modular weight $-1$.
On the other hand, the model in Ref.~\cite{Kobayashi:2019mna} is a global supersymmetric model where the superpotential has the vanishing weight.
Thus, we rearrange modular weights of chiral superfields.
We assign the modular weight $-1$ to all of the left-handed and right-handed 
leptons and Higgs fields.

For the $A_4$ flavor symmetry, the left-handed lepton doublets, $(L_e,L_\mu,L_\tau)^T$ correspond to the $A_4$ triplet $L_{\bf 3}$,
and the right-handed charged leptons are assigned to the $A_4$ singlets of ${\bf 1, 1'', 1'}$, i.e. $e^c_{\mathbf{1}}, \mu^c_{\mathbf{1''}}, \tau^c_{\mathbf{1'}}$;
while the up- and down-sector Higgs fields, $H_u$ and $H_d$, are assigned to the trivial singlet.
The charge assignment of the fields and modular forms is summarized in Table \ref{tb:charge}.
\begin{table}[h]
	\centering
	\begin{tabular}{|c||c|c|c|c|c|c|c|} \hline
		        & $L_{\mathbf{3}}$ & $e^c_{\mathbf{1}},\mu^c_{\mathbf{1''}},\tau^c_{\mathbf{1'}}$ & $H_{u}$ & $H_{d}$ & $Y_{\mathbf{3}}$ & $Y_{\mathbf{1'}}$ & $Y_{\mathbf{1''}}$ \\ \hline \hline \rule[14pt]{0pt}{0pt}
		$SU(2)$ & $2$              & $1$                                                          & $2$     & $2$     & $1$              & $1$               & $1$                \\
		$A_4$   & $3$              & $1,1'',1'$                                                   & $1$     & $1$     & $3$              & $1'$              & $1''$              \\
		$-k_I$  & $-1$             & $-1$                                                         & $-1/2$  & $-1$    & $k = 2$          & $k = 2$           & $k = 2$            \\ \hline
	\end{tabular}
	\caption{
	The charge assignment of $SU(2)$, $A_4$, and the modular weight ($-k_I$ for fields and $k$ for coupling $Y$).}
	\label{tb:charge}
\end{table}

The superpotential of the neutrino mass term is given by the Weinberg operator:
\begin{align}
	\label{eq:wnu}                                                                                          
	W_\nu = \frac{1}{\Lambda}\bigg[ Y_{A_4\mathbf{3}} + a Y_{A_4\mathbf{1''}} + b Y_{A_4\mathbf{1'}} \bigg] 
	L_{\mathbf{3}} L_{\mathbf{3}} H_u H_u,                                                                  
\end{align}
where $\Lambda$ is a cut-off scale; and parameters $a$ and $b$ are complex constants in general.
The superpotential of the mass term of the charged leptons is described as
\begin{align}
	\label{eq:we}                                                                                         
	W_e = \bigg[ \alpha e^c_{\mathbf{1}} + \beta \mu^c_{\mathbf{1''}} + \gamma \tau^c_{\mathbf{1'}}\bigg] 
	Y_{A_4\mathbf{3}} L_{\mathbf{3}}  H_d,                                                                
\end{align}
where $\alpha$, $\beta$ and $\gamma$ are taken to be real and positive without loss of generality.

The superpotential $w$ in the global supersymmetry basis is related to one in the supergravity basis by $|w|^2=e^{K}|W|^2$, i.e. $|w_\nu|^2 = |W_\nu|^2/|{\tau - \bar \tau}|$ and $|w_e|^2 =  |W_e|^2/|{\tau - \bar \tau}|$ \footnote{
	Here, we treat $\tau$ as a vacuum expectation value, but not a holomorphic field.}.
For canonically normalized lepton fields, the Majorana neutrino mass matrix is written as follows:
\begin{align}
	M_\nu = \frac{\langle H_u \rangle ^2}{\Lambda'}\left[
	\begin{pmatrix}
	2Y_3 & -Y_5 & -Y_4 \\
	-Y_5 & 2Y_4 & -Y_3 \\
	-Y_4 & -Y_3 & 2Y_5 
	\end{pmatrix}
	+ a Y_1
	\begin{pmatrix}
	0    & 1    & 0    \\
	1    & 0    & 0    \\
	0    & 0    & 1    
	\end{pmatrix}
	+ b Y_2
	\begin{pmatrix}
	0    & 0    & 1    \\
	0    & 1    & 0    \\
	1    & 0    & 0    
	\end{pmatrix}  \right],
\end{align}
where 
\begin{equation}
	\Lambda' = \Lambda |\tau - \bar \tau|^{3/2},
\end{equation}
while the charged lepton matrix is given as 
\begin{align}
	M_e = \langle H_d \rangle
	\begin{pmatrix}
	\alpha' & 0      & 0       \\
	0       & \beta' & 0       \\
	0       & 0      & \gamma' 
	\end{pmatrix}
	\begin{pmatrix}
	Y_3     & Y_5    & Y_4     \\
	Y_4     & Y_3    & Y_5     \\
	Y_5     & Y_4    & Y_3     
	\end{pmatrix}_{RL},
	\label{lepton}
\end{align}
with 
\begin{equation}
	\alpha' = \alpha |\tau - \bar \tau|^{1/2}, \qquad \beta' = \beta |\tau - \bar \tau|^{1/2}, 
	\qquad \gamma' = \gamma |\tau - \bar \tau|^{1/2}.
\end{equation}
The parameters $\alpha', \beta', \gamma'$ are determined by the observed charged lepton masses and the value of  $\tau$.

\begin{figure}[h]
	\centering
	\includegraphics[bb=0 0 850 650,width=0.6\linewidth]{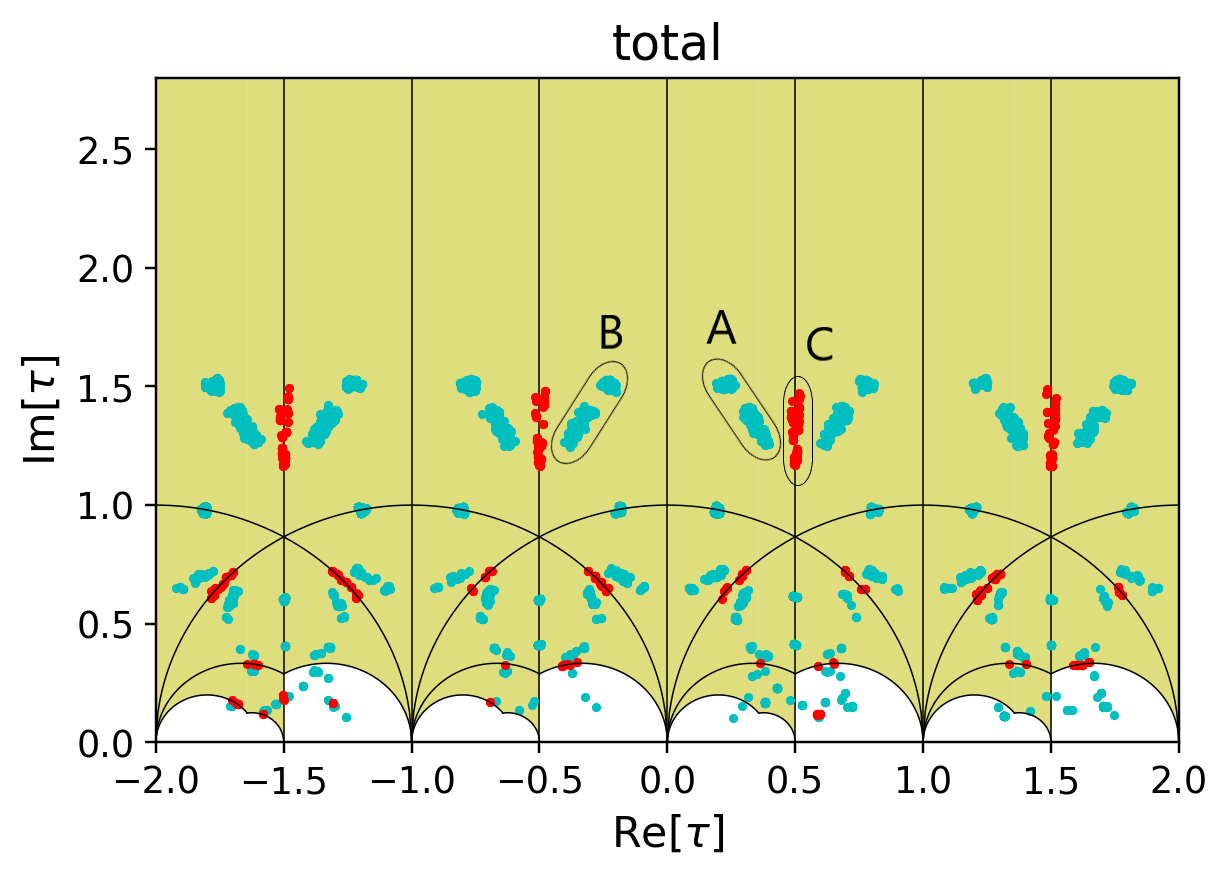}
	\caption{ Allowed regions of $\tau$ on the ${\rm Re}[\tau]$--$ {\rm Im}[\tau]$ plane.
		The fundamental domain of $\Gamma(4)$ is shown by olive-green.
	Cyan-points and red-points denote the cases of NH and IH, respectively.}
	\label{fig:tau}
\end{figure}

We take $a$ and $b$ to be real in order to present a simple viable model.
We scan parameters in the following ranges:
\begin{align}
	\tau = [-2.0, 2.0] + i\  [0.1, 2.8], \quad 
	a = [-15, 15], \quad                       
	b = [-15, 15],                             
\end{align}
where the fundamental domain of $\Gamma(4)$ is taken into account.
The lower-cut $0.1$ of ${\rm Im[\tau]}$ is artificial to keep the accurate numerical calculation.
The upper-cut $2.8$ is large enough to estimate the modular forms.
We input the experimental data within $3\,\sigma$ C.L.\cite{NuFIT} of three mixing angles in the lepton mixing matrix \cite{Tanabashi:2018oca} in order to constrain the magnitudes of parameters.
We also put the observed neutrino mass ratio $\Delta m_{\rm sol}^2/\Delta m_{\rm atm}^2$ and the cosmological bound for the neutrino masses $\sum m_i < 0.12$ [eV]~\cite{Vagnozzi:2017ovm,Aghanim:2018eyx}.
There are two possible spectra of neutrinos masses $m_i$, which are the normal hierarchy (NH), $m_3>m_2>m_1$, and the inverted hierarchy (IH), $m_2>m_1>m_3$.
Figure \ref{fig:tau} shows allowed regions for   NH (Cyan) and IH(Red), respectively.

\section{Modulus stabilization}

We study the modulus stabilization in the $A_4$ symmetric model where the $S_4$ modular symmetry is assumed to be broken by anomalies.
For the modulus stabilization, we need a modulus-dependent superpotential $W(\tau)$ which may be induced by non-perturbative effects.
Such superpotential $W(\tau)$ must have the modular weight $-1$ for the modular invariance.
However, there is no modular form of odd weights for $\Gamma(4)$.
We need some mechanism to generate the superpotential term for modulus stabilization.

Here, we assume that the following superpotential,
\begin{equation}
	\label{eq:W-tau}
	W = \Lambda_d^{(3)}  (Y^{(4)}(\tau))^{-1},
\end{equation}
where we assumed that $\Lambda_d^{(3)}$ has the modular weight 3.
This modulus superpotential may be induced from  the condensation $\langle Q \bar Q \rangle \neq 0$ 
in the hidden sector by strong dynamics such as supersymmetric QCD, 
and $\Lambda_d^{(3)}$ is the  dynamical scale which is related to the condensation, e.g. $\Lambda_d^{(3)} = m\langle Q \bar Q \rangle$
(See for Appendix C.).
We assume the above superpotential from the bottom-up viewpoint.

\begin{figure}[h!]
	\centering
	\includegraphics[bb=0 0 400 280,width=0.6\linewidth]{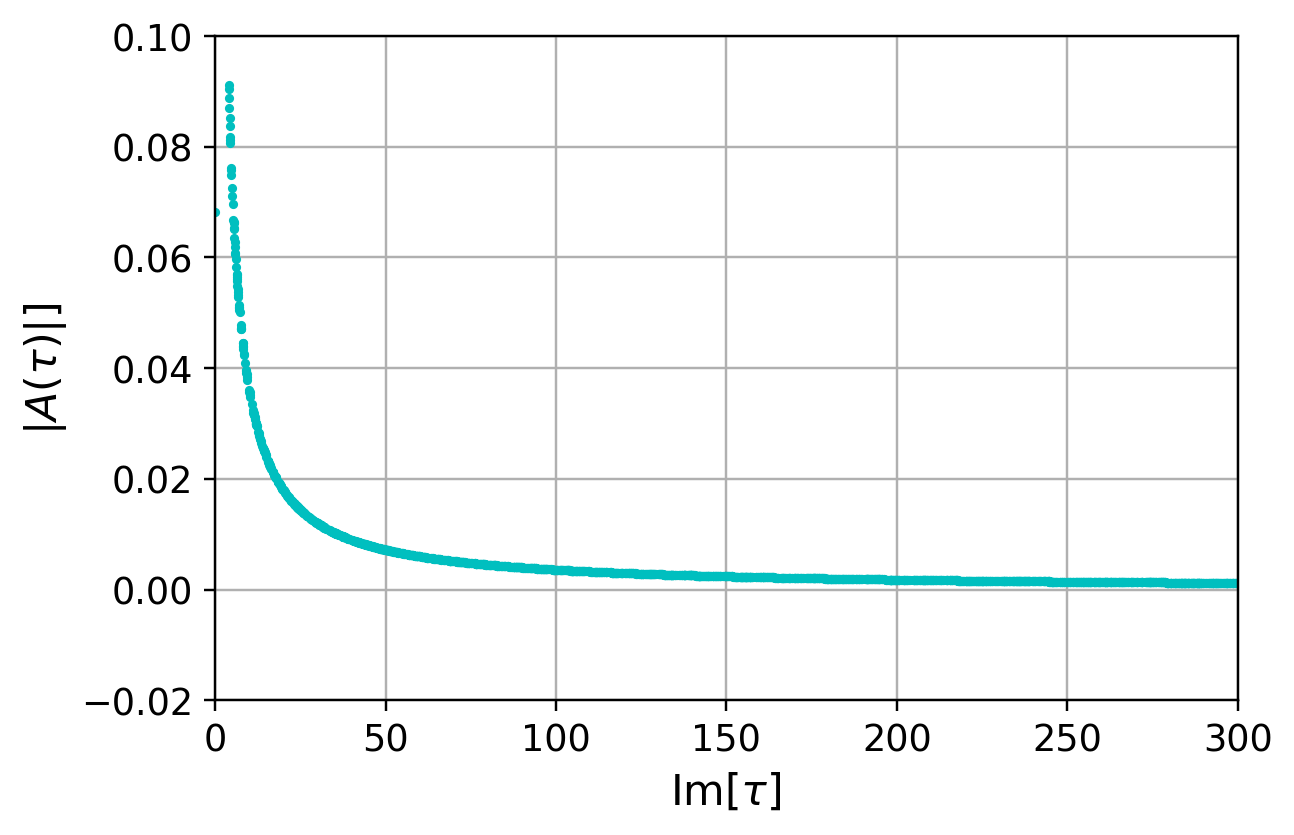}
	\caption{$|A(\tau)| \equiv |D_\tau W|/\Lambda_d^{(3)}$ at the slice ${\rm Re}[\tau] =0$. }
	\label{fig:running}
\end{figure}

The scalar potential in supergravity theory is written by using $K$ in Eq.(\ref{eq:Kahler-tau}) and $W$ in Eq.(\ref{eq:W-tau}) as
\begin{equation}
	V=e^K \left( (K_{\tau \bar \tau}^{-1}|D_\tau W|^2 -3|W|^2 \right),
\end{equation}
where
\begin{equation}
	\label{eq:DW}
	D_\tau W = K_\tau W + W_\tau,
\end{equation}
with $K_\tau = \partial K/\partial \tau$ and $W_\tau = \partial W/ \partial \tau$.
We analyze the minimum of the above scalar potential $V$ by examining the stationary condition, $\partial V/\partial \tau =0$.
If there is a solution in the following equation,
\begin{equation}\label{eq:DW=0}
	D_\tau W = 0,
\end{equation}
we have $\partial V/\partial \tau =0$.
Such a solution is a candidate for the potential minimum
and corresponds to a supersymmetric minimum.
However, the above scalar potential has no proper supersymmetric minimum.
For the slice of ${\rm Re}(\tau) =0$, the value of $|A(\tau)| \equiv |D_\tau W|/\Lambda_d^{(3)}$ is shown in Figure \ref{fig:running} for larger values of ${\rm Im}[\tau]$. 
The value $|D_\tau W|$ becomes to vanish as ${\rm Im}[\tau] \rightarrow \infty$.
Similarly,  $|D_\tau W|$ becomes to vanish as ${\rm Im}[\tau] \rightarrow 0$, because ${\rm Im}[\tau] \rightarrow 0$ and $\infty$ are related to each other by the $S$ transformation.
The minimum corresponds to ${\rm Im}[\tau] \rightarrow  0$ and $\infty$.
There is no supersymmetric minimum for a finite value of $\tau$.

\begin{figure}[h!]
	\centering
	\includegraphics[bb=0 0 400 280,width=0.6\linewidth]{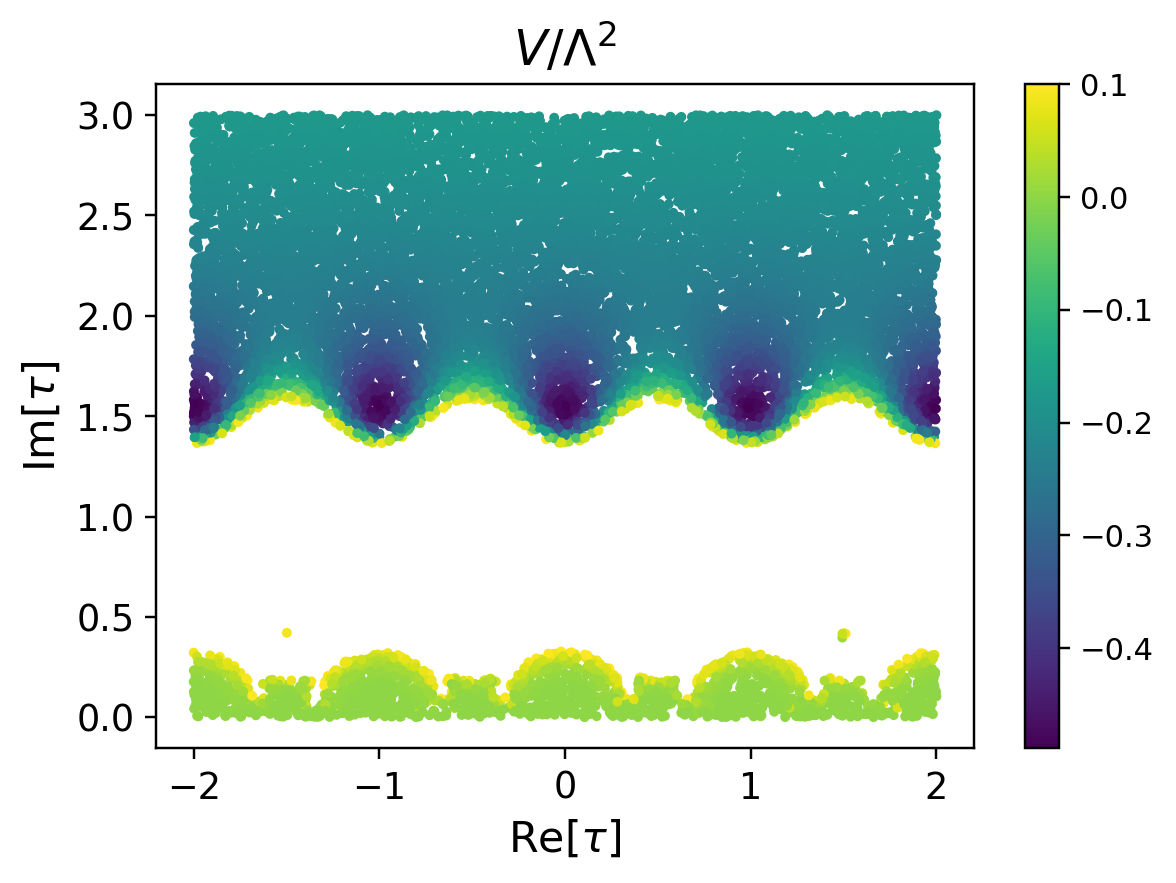}
	\caption{A contour map of the scalar potential for $W$ in Eq.~(\ref{eq:W-tau}). The potential minima 
		correspond to $\tau = 1.54i +n$, where $n$ is integer. }
	\label{fig:V-minima}
\end{figure}

On the other hand, the scalar potential has non-supersymmetric minima as shown in Figure \ref{fig:V-minima}.
The minima correspond to $\tau = 1.54i +n$, where $n$ is integer.
Unfortunately, these minima do not lead to realistic lepton mass matrices. (See Figure \ref{fig:tau}.)
We have $V \sim - 0.5 \times (\Lambda_d^{(3)})^2$, and the modulus mass squared $m^2_\tau \sim 100 \times (\Lambda_d^{(3)})^2$ at these minima.
We need to uplift the vacuum energy by other supersymmetry breaking effects in order to realize almost vanishing vacuum energy $V \approx 0$.
Such uplifting effects may not shift significantly the stabilized value $\tau = 1.54i +n$ because the modulus mass squared is large compared with the negative vacuum energy $V \sim - 0.5 \times (\Lambda_d^{(3)})^2$.

\begin{figure}[h!]
	\centering
	\includegraphics[bb=0 0 400 280,width=0.6\linewidth]{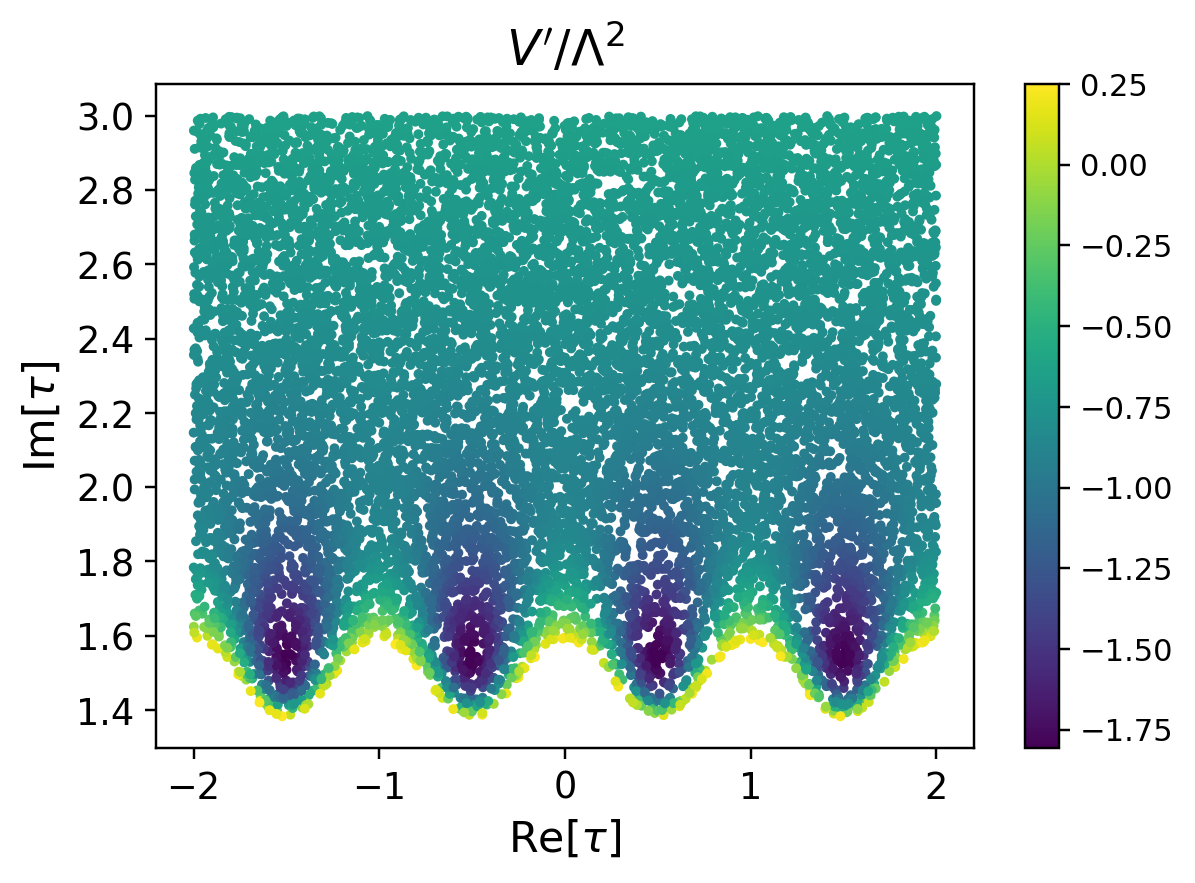}
	\caption{ A contour map of the scalar potential for $W$ in Eq.(\ref{eq:W-tau-2}). The potential minima 
		correspond to $\tau = 1.55i +n/2$, where $n$ is odd. }
	\label{fig:V'-minima}
\end{figure}

Alternatively, we assume the following superpotential,
\begin{equation}\label{eq:W-tau-2}
	W = \Lambda_d^{(-5)}  Y^{(4)}(\tau) ,
\end{equation}
where we assumed that  $\Lambda_d^{(-5)}$ has the modular weight $-5$. 
However, the corresponding scalar potential has no proper supersymmetric minimum.
Fig.~\ref{fig:V'-minima}  shows the corresponding scalar potential.
Its minima correspond to $\tau = 1.55i + n/2$, where $n$ is odd.
Unfortunately, these values also do not lead to realistic lepton mass matrices.(See Figure \ref{fig:tau}.)
We have $V \sim - 2 \times (\Lambda_d^{(-5)})^2$, and the modulus mass squared $m^2_\tau \sim 400 \times (\Lambda_d^{(-5)})^2$ at these minima.
The effects from uplifting the vacuum energy to $V \approx 0$ on the stabilized value $\tau = 1.55i +n$ is small because the modulus mass squared is large compared with the negative vacuum energy $V \sim - 2 \times (\Lambda_d^{(-5)})^2$.

We can use the modular form $Y^{(6)}(\tau)$ instead of $Y^{(4)}(\tau)$ in 
Eqs.~(\ref{eq:W-tau})  and (\ref{eq:W-tau-2}).
When we replace  $Y^{(4)}(\tau)$ in Eq.~(\ref{eq:W-tau})  by  $Y^{(6)}(\tau)$, 
the corresponding scalar potential has the minimum at $\tau = 1.68i + 1/2$.
On the other hand, when we replace  $Y^{(4)}(\tau)$ in Eq.~(\ref{eq:W-tau-2})  by  $Y^{(6)}(\tau)$, 
the corresponding scalar potential has the minimum at $\tau = 1.69i $.
Unfortunately, these values of $ \tau$ are not proper to realize the lepton masses and mixing angles.

Thus, we can stabilize the modulus, but its values are not realistic when 
the superpotential includes a single modular form.
We need more terms to stabilize the modulus at a proper value.
For example, we assume the following superpotential,
\begin{equation}
	\label{eq:W-tau-rho}
	W = \Lambda_d^{(3)}  (Y^{(4)}(\tau))^{-1} +\Lambda_d^{(5)} (Y^{(6)}(\tau))^{-1} ,
\end{equation}
where $\Lambda_d^{(5)}$ is assumed to have the modular weight $5$.
Here, we define $\rho = \Lambda_d^{(5)} / \Lambda_d^{(3)}$.
This superpotential  always has a supersymmetric minimum for a finite value of $\rho$.
We focus on such a supersymmetric minimum.

\begin{figure}[h!]
	\begin{tabular}{ccc}
		\begin{minipage}{0.475\linewidth}                                                                         
		\includegraphics[bb=0 0 400 280,width=\linewidth]{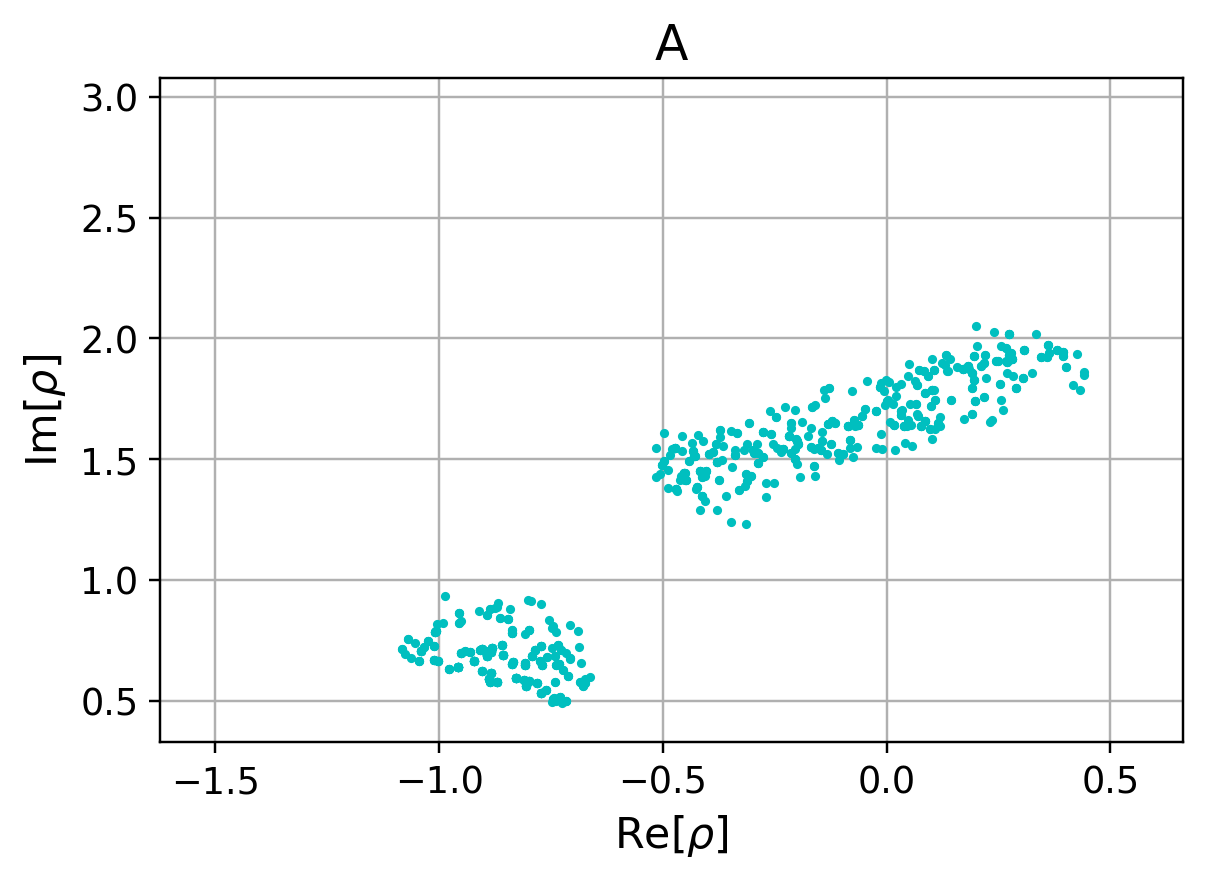}                                      
		\caption{Values of $\rho$ corresponding to $\tau$ in the A region for $ W$ in Eq.~(\ref{eq:W-tau-rho}).}  
		\label{fig:rho-A}                                                                                         
		\end{minipage}                                                                                            
		\phantom{=}                                                                                               
		\begin{minipage}{0.475\hsize}                                                                             
		\includegraphics[bb=0 0 400 280,width=\linewidth]{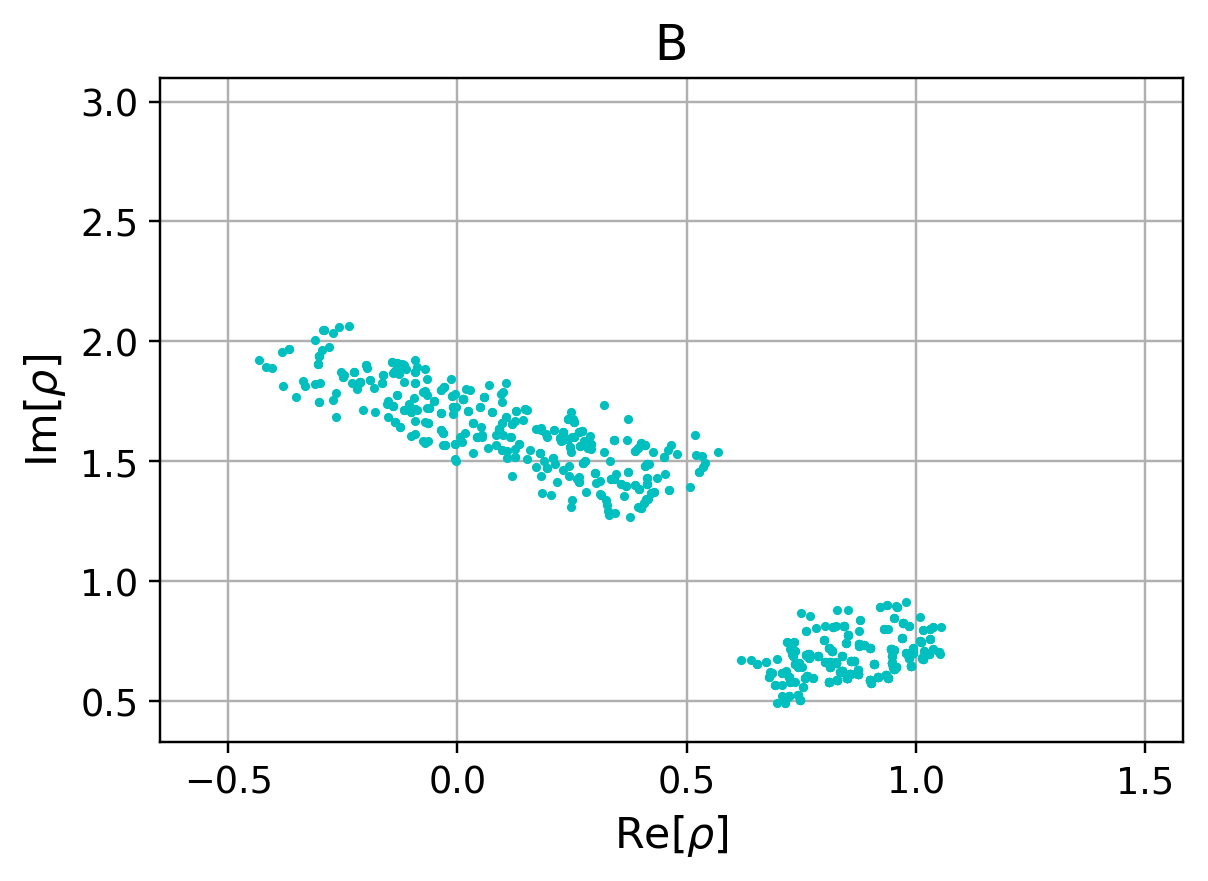}                                      
		\caption{ Values of $\rho$ corresponding to $\tau$ in the B region for $ W$ in Eq.~(\ref{eq:W-tau-rho}).} 
		\label{fig:rho-B}                                                                                         
		\end{minipage}                                                                                            
	\end{tabular}
\end{figure}

\begin{figure}[h!]
	\centering
	\includegraphics[bb=0 0 400 300,width=0.475\linewidth]{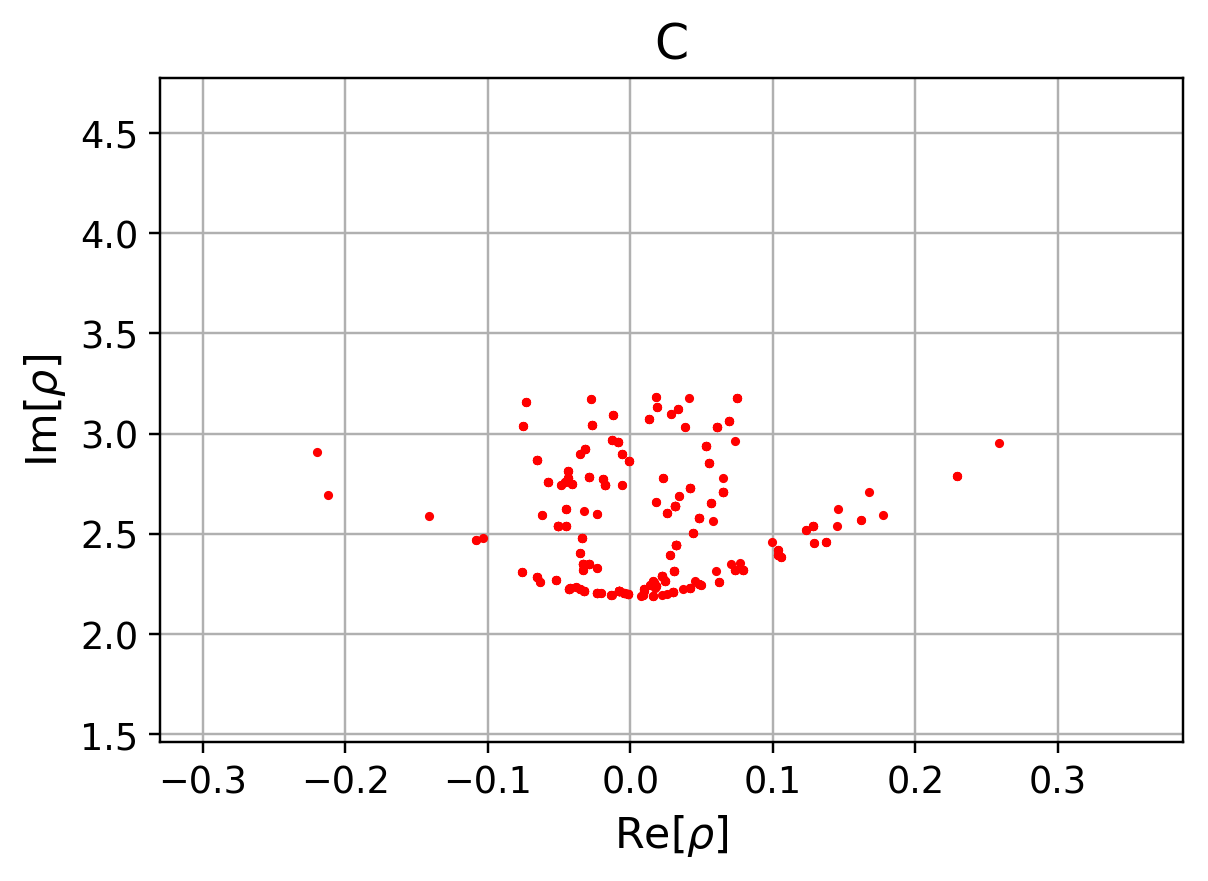}
	\caption{Values of $\rho$ corresponding to $\tau$ in the C region $ W$ in Eq.~(\ref{eq:W-tau-rho}).}
	\label{fig:rho-C}
\end{figure}

For smaller values of $\tau$, the K\"ahler potential of Eq.(\ref{eq:Kahler-tau}) may have corrections.
Thus, we restrict ourselves to the case with $\tau = {\cal O}(1)$.
That is, we study the A, B and C regions in Figure \ref{fig:tau}.
We can choose a proper value of $\rho$ such that $\tau$ is fixed to be a value in the A, B and C regions through Eq.~(\ref{eq:DW=0}).
Figures \ref{fig:rho-A}, \ref{fig:rho-B}, and \ref{fig:rho-C} show the values of $\rho$ obtained from each value of $\tau$ in the A, B and C regions.
The values of $\tau$ in the A, B, C regions are obtained by 
$|\rho|={\cal O}(1)$.
That is, the $Y^{(4)}$ contribution is comparable with the $Y^{(6)}$ contribution.
The A and   B regions are almost symmetric  to ${\rm Im}[\rho]$ axis.
Furthermore, the values of $\rho$ for the region $C$ are imaginary dominant.
Hence, the A, B, and C regions are realized by  different values of $\rho$.
At any rate,  both $Y^{(4)}(\tau)$ and $Y^{(6)}(\tau) $ are important to fix favorable values of $\tau$ in the potential.
The IH mass spectrum can be realized only in the C region, that is, ${\rm Re}[\rho] \approx [-0.25, 0.25]$, 
and ${\rm Im}[\rho] \approx [2.2,3.2] $.

At these minima, we obtain typical values of $|W_{\tau \tau}|={\cal O}(10) \times \Lambda_d^{(-3)}$ in the A and B regions, while 
in the C region we can obtain larger  $|W_{\tau \tau}|={\cal O}(100) \times \Lambda_d^{(-3)}$.
Thus, the modulus mass is estimated $m_\tau ={\cal O}(10-100) \Lambda_d^{(-3)}$ in the unit of $M_P=1$.
These minima correspond to the anti-de Sitter supersymmetric vacua whose negative vacuum energy is written by $V = -3e^K|W|^2 = -3 |W|^2/|\tau -\bar \tau|$.
Here, $|W|/\Lambda_d^{(-3)}={\cal O}(1)$ in all of the A, B and C regions.
Thus, the gravitino mass $m_{3/2}$ is estimated by $m_{3/2} = {\cal O}(1) \times \Lambda_d^{(-3)}$ in the unit of $M_P= 1$.
We need to uplift the vacuum energy to realize almost vanishing vacuum energy, $V \approx 0$ by supersymmetry breaking.
Uplifting may shift stabilized values of $\tau$, but such a shift  $\delta \tau$  is very small because we can estimate $\delta \tau/\tau \sim m^2_{3/2}/m^2_\tau = {\cal O}(10^{-4}-10^{-2})$.


\begin{figure}[h!]
	\begin{tabular}{ccc}
		\begin{minipage}{0.475\linewidth}                                                                  
		\includegraphics[bb=0 0 400 280,width=\linewidth]{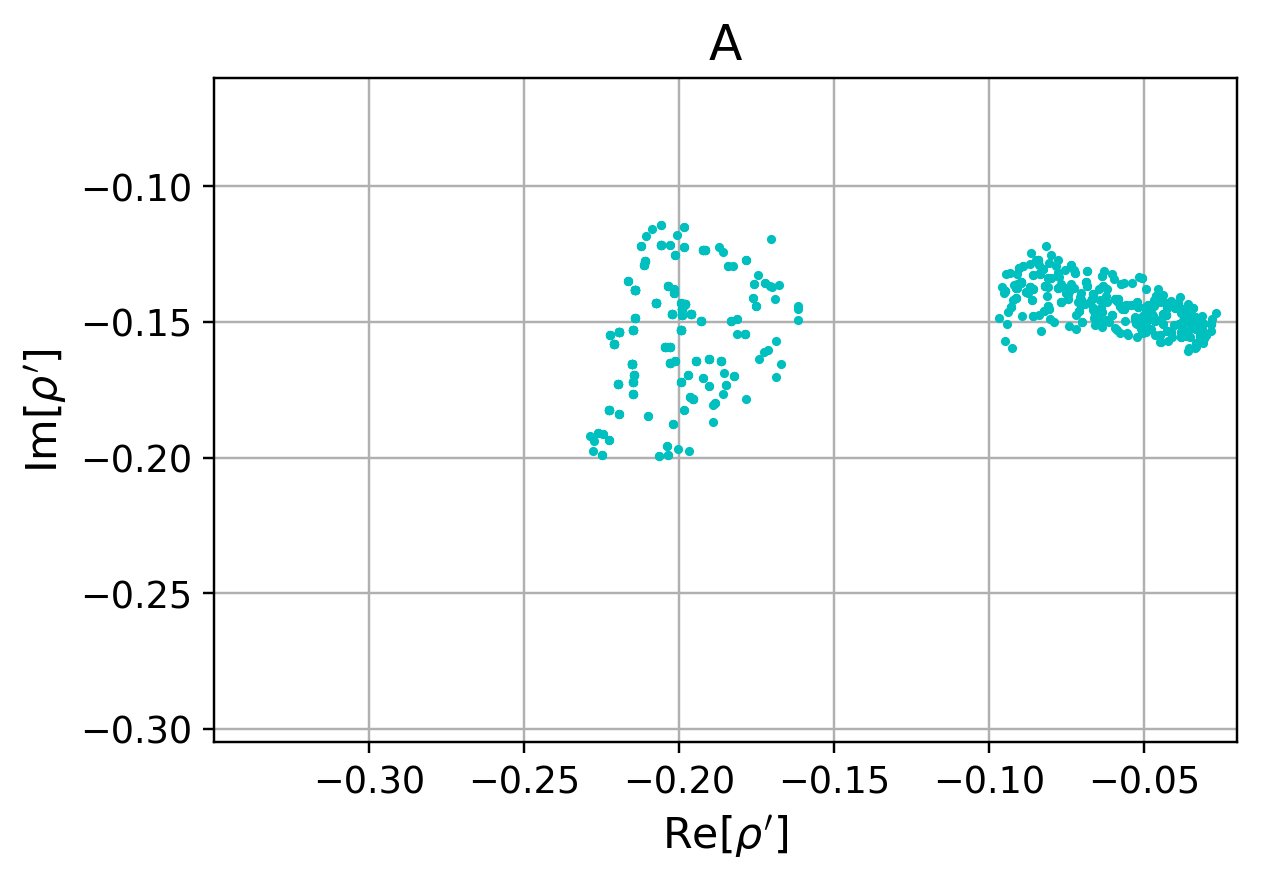}                             
		\caption{Values of $\rho'$ corresponding to $\tau$ in the A region for $W$ in Eq.~(\ref{eq:W'}).}  
		\label{fig:rho_a-A}                                                                                
		\end{minipage}                                                                                     
		\phantom{=}                                                                                        
		\begin{minipage}{0.475\hsize}                                                                      
		\includegraphics[bb=0 0 400 280,width=\linewidth]{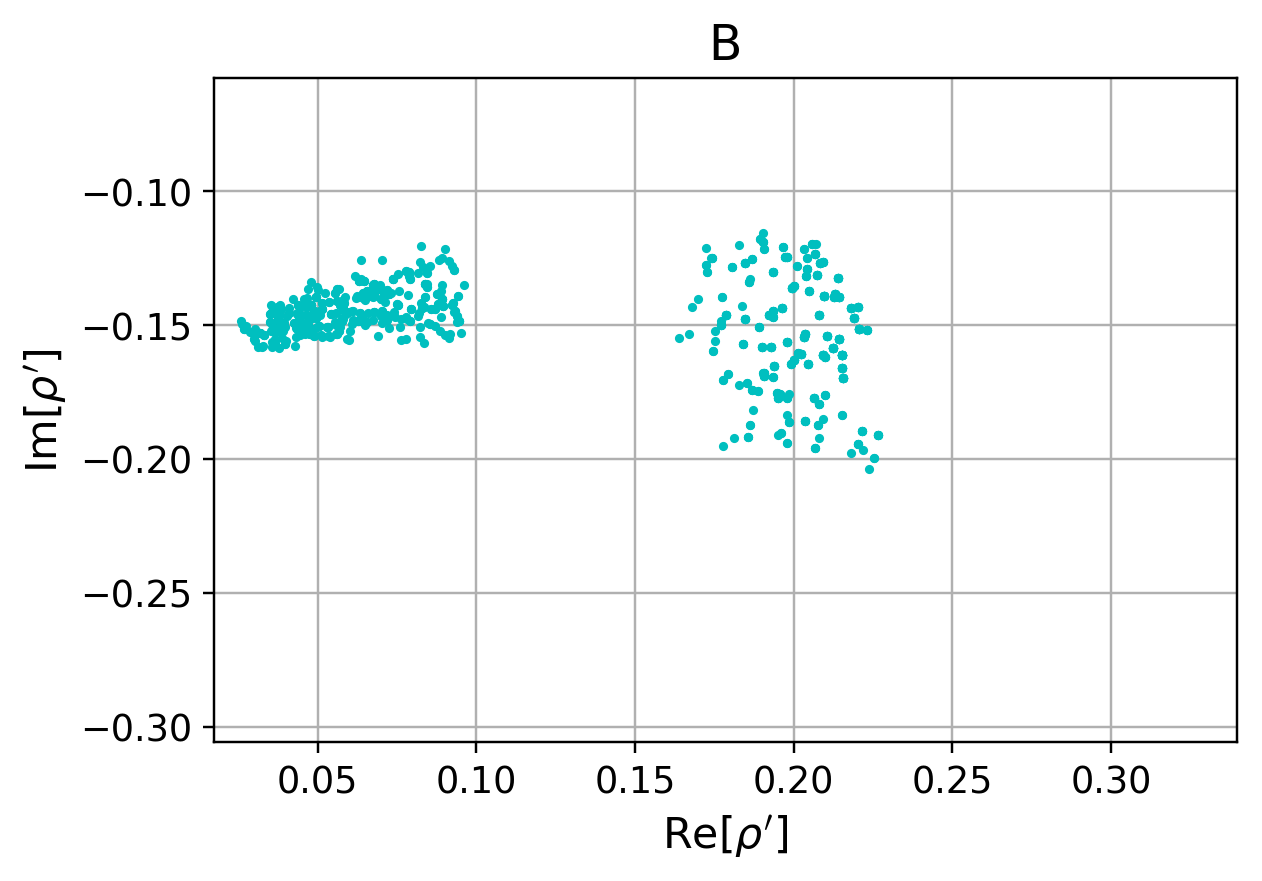}                             
		\caption{ Values of $\rho'$ corresponding to $\tau$ in the B region for $W$ in Eq.~(\ref{eq:W'}).} 
		\label{fig:rho_a-B}                                                                                
		\end{minipage}                                                                                     
	\end{tabular}
\end{figure}

\begin{figure}[h!]
	\centering
	\includegraphics[bb=0 0 400 300,width=0.5\linewidth]{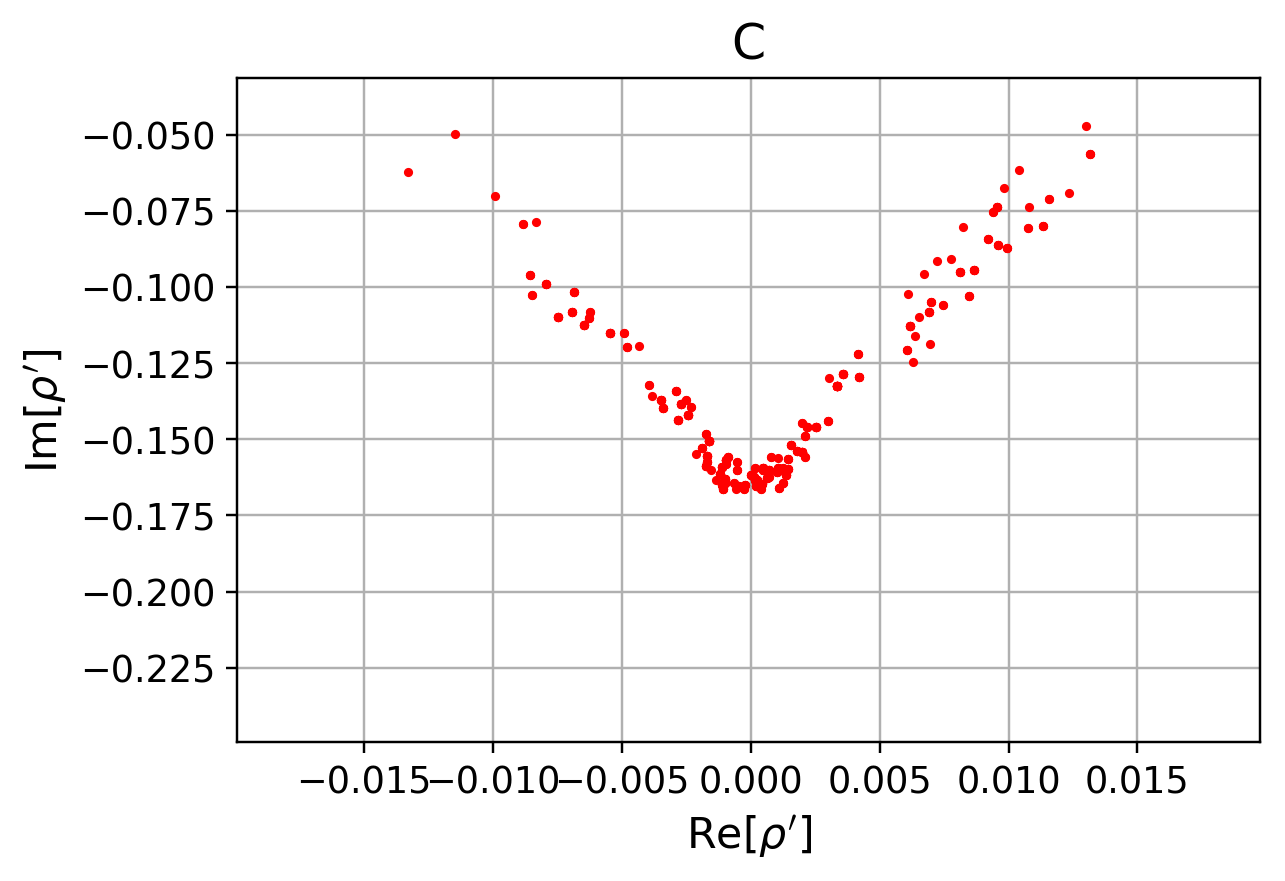}
	\caption{Values of $\rho'$ corresponding to $\tau$ in the C region $W$ in Eq.~(\ref{eq:W'}).}
	\label{fig:rho_a-C}
\end{figure}

Similarly, we can use the following superpotential:
\begin{equation}
	\label{eq:W'}
	W = \Lambda_d^{(-5)}  Y^{(4)}(\tau) +\Lambda^{(-7)} Y^{(6)}(\tau) ,
\end{equation}
by assuming that non-perturbative effects generate it and $\Lambda_d^{(-5)}$ and  $\Lambda_d^{(-7)}$ have 
the modular weights $-5$ and $-7$.
Here, we define $\rho' = \Lambda_d^{(-7)}/ \Lambda_d^{(-5)}$.
Then, similarly we can study the modulus stabilization by using this superpotential.
Again, we analyze the supersymmetric condition Eq.~(\ref{eq:DW=0}).
We can find values of the modulus $\tau$, which satisfy the supersymmetric condition Eq.~(\ref{eq:DW=0}), by choosing a proper value of $\rho'$.
Figures \ref{fig:rho_a-A}, \ref{fig:rho_a-B}, and  \ref{fig:rho_a-C} show such values of $\rho'$ leading to the values of $\tau$ in the  A, B and C regions.
At these minima, we obtain typical values of $|W_{\tau \tau}|={\cal O}(10) \Lambda_d^{(-5)}$ in the A and B regions, while 
in the C region we obtain $|W_{\tau \tau}|={\cal O}(10^4) \Lambda_d^{(-5)}$.
Therefore, the modulus mass is estimated $m_\tau ={\cal O}(10) \Lambda_d^{(-5)}$ in the A and B regions, while the modulus mass 
can be larger in the C region such as $m_\tau ={\cal O}(10^4) \Lambda_d^{(-5)}$.
These minima correspond to the anti-de Sitter supersymmetric vacua whose negative vacuum energy is written by  $V = -3e^K|W|^2 = -3 |W|^2/|\tau -\bar \tau|'$, 
where $|W|/\Lambda_d^{(-5)}={\cal O}(1)$ in the A and B regions and $|W|/\Lambda_d^{(-5)}={\cal O}(10)$ in the C region.
The gravitino mass $m_{3/2}$ is estimated by $m_{3/2} = {\cal O}(1)\Lambda_d^{(-5)}$ in the A and B regions, and 
$m_{3/2} = {\cal O}(10)\Lambda_d^{(-5)}$ in the C region.
Thus, the shift $\delta \tau$ by uplifting will be small.


As results, we can stabilize the modulus $\tau$ at realistic values in three regions $A, B, C$ by using 
both the superpotential terms, (\ref{eq:W-tau-rho}) and (\ref{eq:W'}) with proper values of 
the parameters, $\rho$ and $\rho'$.
In the next section, we study phenomenological aspects of these three regions following 
the modulus stabilization by both the superpotential terms, (\ref{eq:W-tau-rho}) and (\ref{eq:W'}).

\section{Phenomenological aspects of leptons}

In this section, we discuss phenomenological  results 
derived from the mass matrices of charged leptons and neutrinos for three regions $A, B, C$ of the modulus in Fig.\ref{fig:tau}, respectively.

\subsection{Region of $A$ }
Let us present numerical results in the region $A$ of the modulus $\tau$.
The parameter $\rho$ which realizes the potential minimum for the superpotential (34) is shown in the ${\rm Re}[\rho]$--$ {\rm Im}[\rho]$ plane of Fig.\,\ref{fig:rho-A}, while Fig.\,\ref{fig:rho_a-A} shows $\rho'$ for the potential minimum for the superpotential (35).
In this case, NH is only available. 

\begin{figure}[h!]
	\begin{tabular}{ccc}
		\begin{minipage}{0.475\hsize}                                       
		\includegraphics[bb=0 0 400 280,width=\linewidth]{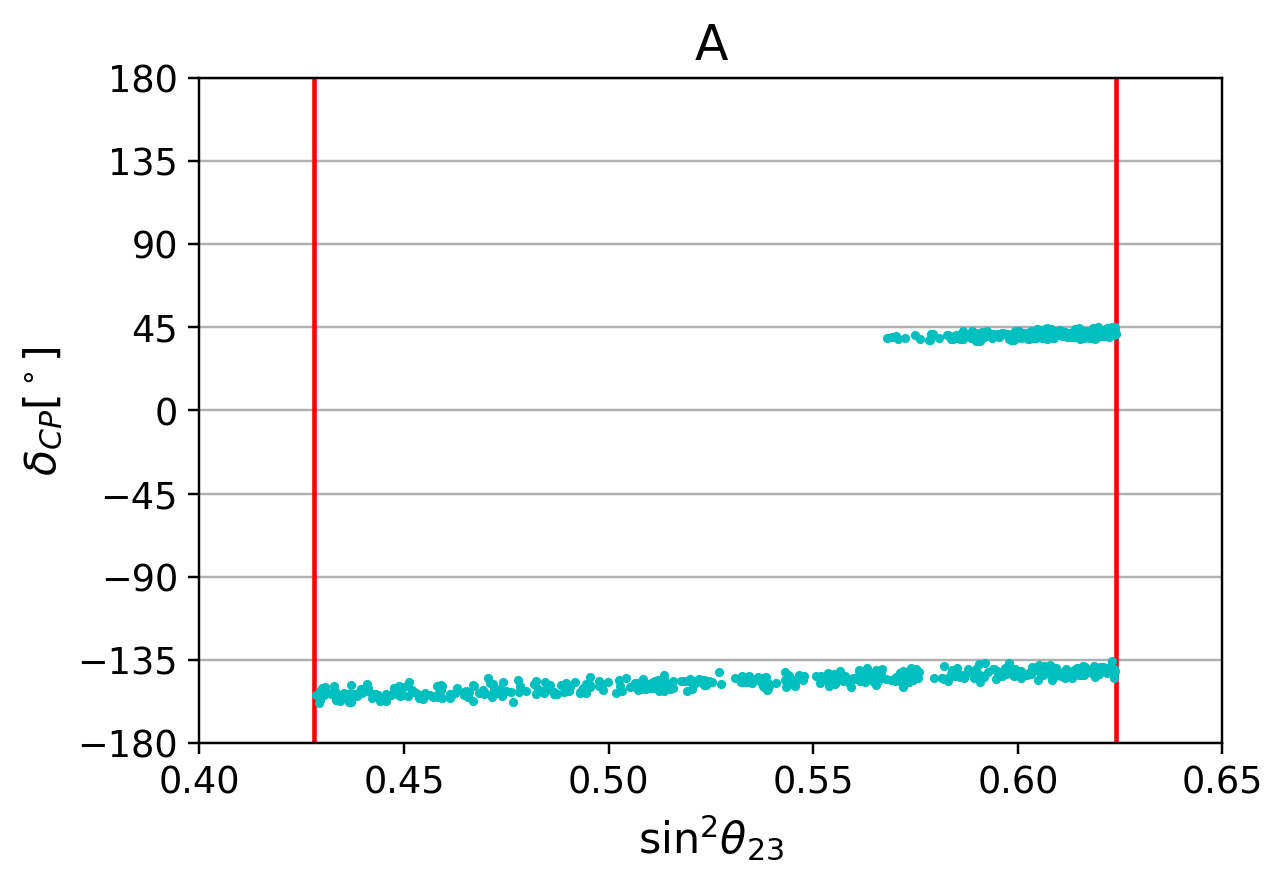}   
		\caption{The predicted region on $\sin^2\theta_{23}$                
		--$\delta_{CP}$ plane in A for NH.                                  
		Vertical red lines denote $3\sigma$ bound of observed data.}        
		\label{fig:delta-23_A}                                              
		\end{minipage}                                                      
		\phantom{=}                                                         
		\begin{minipage}{0.475\hsize}                                       
		\includegraphics[bb=0 0 400 280,width=\linewidth]{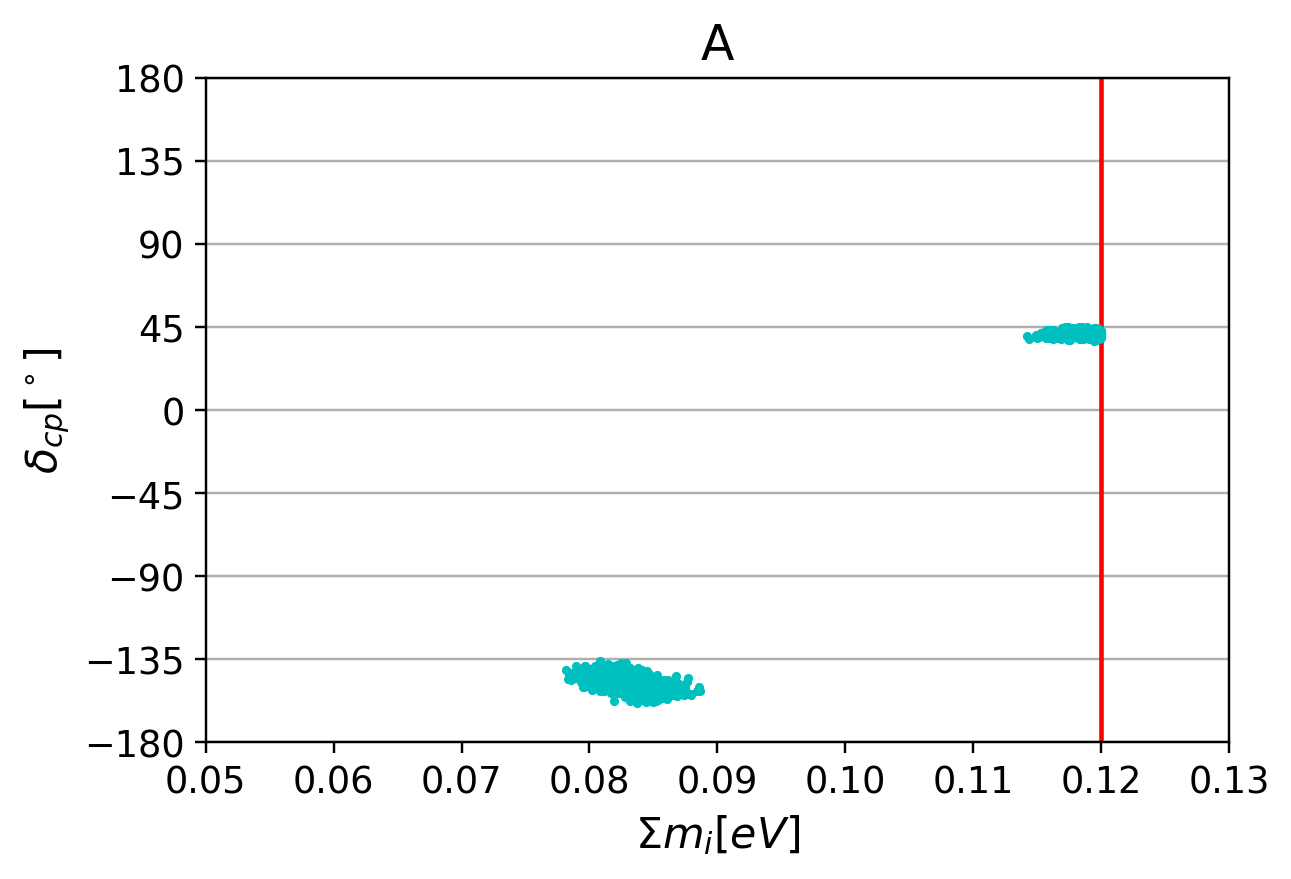} 
		\caption{The sum of neutrino masses $\sum m_i$                      
		dependence of $\delta_{CP}$ in A for NH.                            
		A vertical red line denote the cosmological bound.}                 
		\label{fig:mass-delta_A}                                            
		\end{minipage}                                                      
	\end{tabular}
\end{figure}

At first, we show a correlation between $\delta_{CP}$ and $\sin^2\theta_{23}$ in Fig.\,\ref{fig:delta-23_A}. 
The predicted ranges of $\delta_{CP}$ are  $[-160^\circ,-140^\circ]$ for $\sin^2\theta_{23}\leq 0.57$
and  $[-140^\circ,-135^\circ]$,  $[40^\circ,45^\circ]$ for $\sin^2\theta_{23}\geq 0.57$.
We  present the predicted $\delta_{CP}$ versus the sum of neutrino masses $\sum m_i$ in Fig.\,\ref{fig:mass-delta_A}, where the cosmological bound  $\sum m_i < 120$ [meV] is imposed.
The predicted $\delta_{CP}$ clearly depends on the sum of neutrino masses. For $\sum m_i= [78, 88]$ [meV], $\delta_{CP}$ is predicted in
$[-160^\circ,-135^\circ]$.
Near the cosmological bound,  $\sum m_i= [115, 120]$ [meV], $\delta_{CP}$ is  $[40^\circ,45^\circ]$.

\begin{figure}[h!]
	\begin{tabular}{ccc}
		\begin{minipage}{0.475\hsize}                                     
		\includegraphics[bb=0 0 400 280,width=\linewidth]{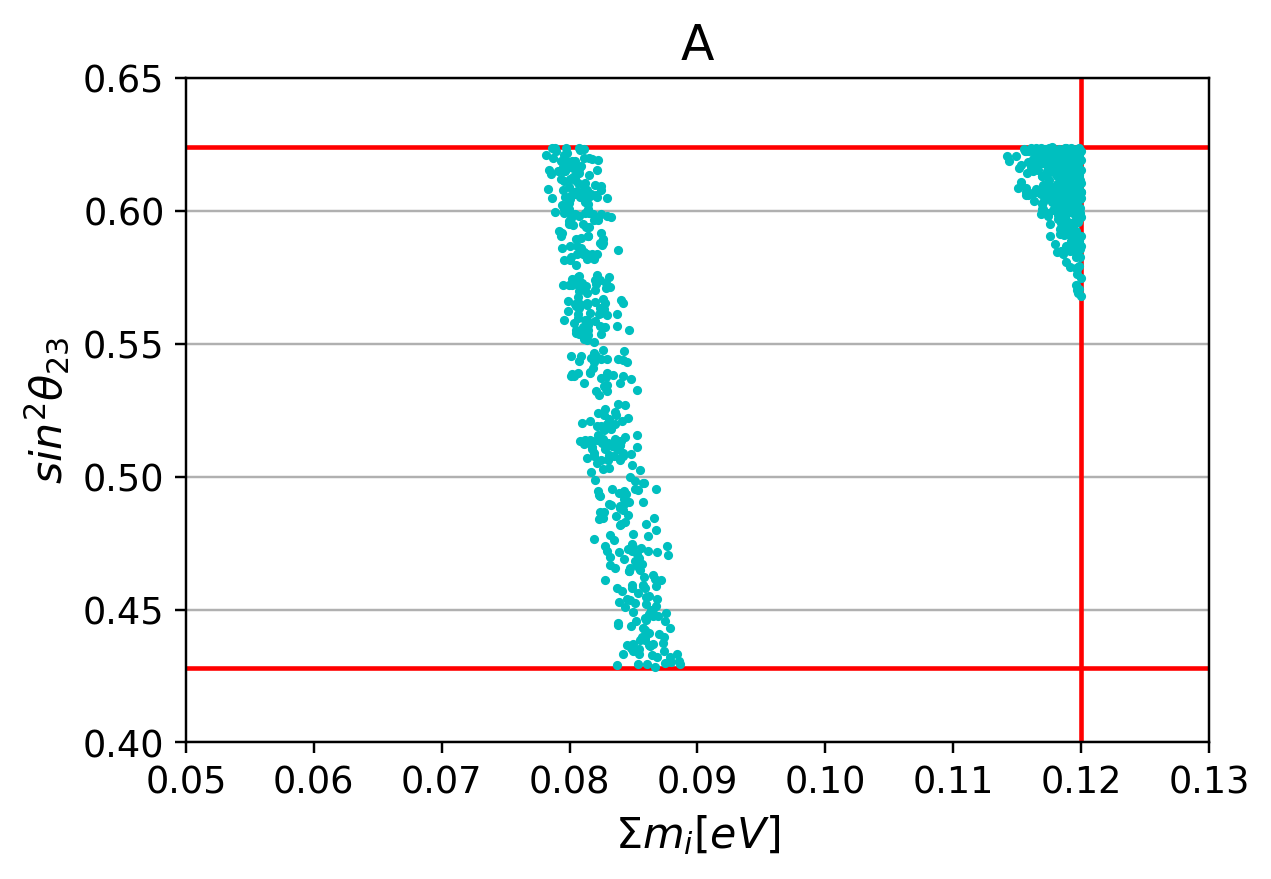}  
		\caption{The sum of neutrino masses $\sum m_i$                    
		dependence of $\sin^2\theta_{23}$ in A for NH. 	}                 
		\label{fig:mass-23_A}                                             
		\end{minipage}                                                    
		\phantom{=}                                                       
		\begin{minipage}{0.475\hsize}                                     
		\includegraphics[bb=0 0 400 280,width=\linewidth]{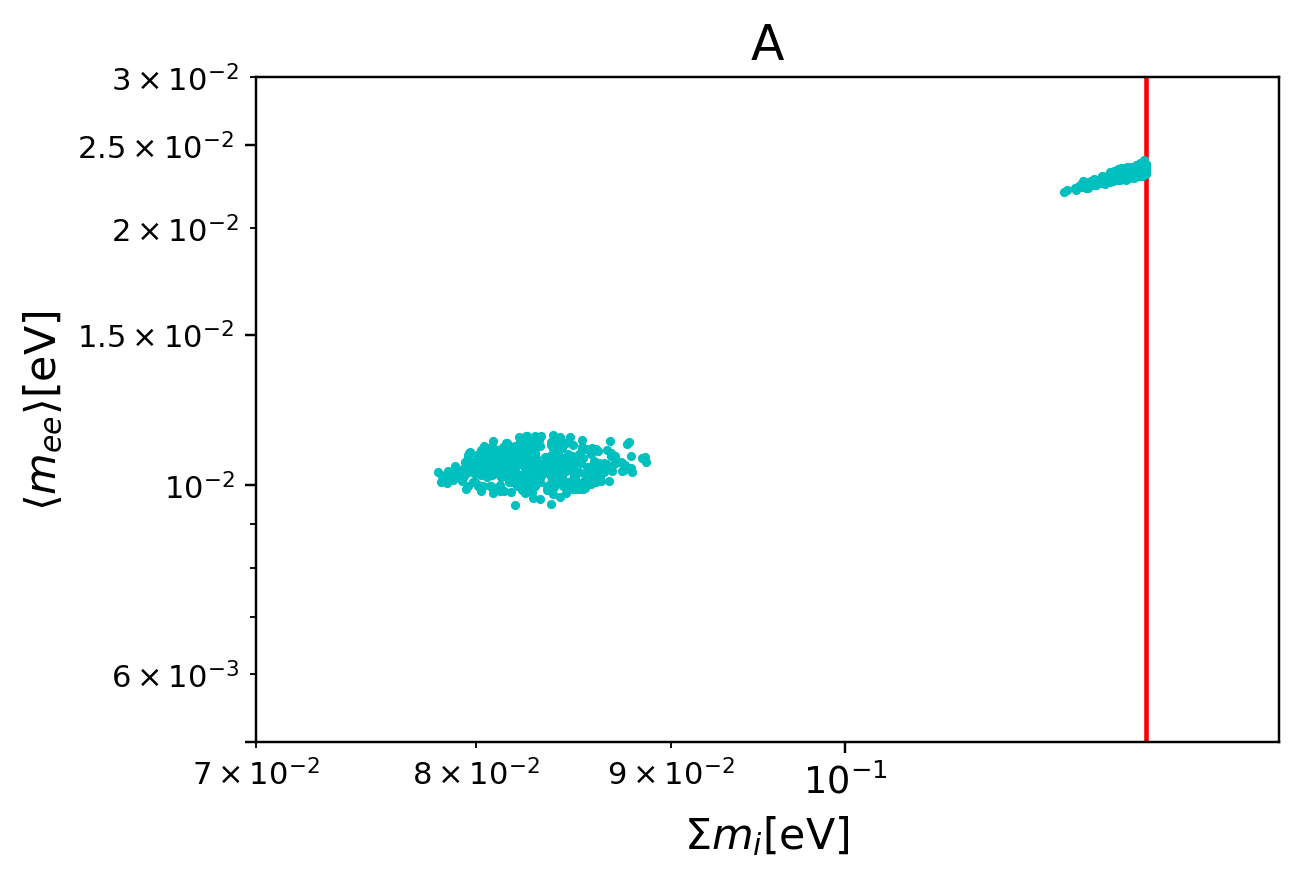} 
		\caption{The predicted  effective neutrino mass                   
		$\langle m_{ee}\rangle$ versus $\sum m_i$  in A for NH. }         
		\label{fig:mass-mee_A}                                            
		\end{minipage}                                                    
	\end{tabular}
\end{figure}

On the other hand, there is no clear neutrino mass dependence for $\sin^2\theta_{23}$ as seen in Fig. \ref{fig:mass-23_A}.
Near the upper bound of  $\sum m_i = 120$ [meV], $\theta_{23}$ is predicted in the second octant.
The effective mass of the $0\nu\beta\beta$ decay $\langle m_{ee}\rangle$  is presented in Fig.\,\ref{fig:mass-mee_A}. The prediction is in the range of $9.4$--$12$ [meV] and $22$--$24$ [meV].
We summarize the prediction for $\langle m_{ee}\rangle$ and $\sum m_i$ in Table \ref{tb:ABC}.


\subsection{Region of $B$}

We discuss numerical results in the region $B$ of  the modulus $\tau$.
The parameter $\rho$ which realizes the potential minimum for the superpotential (34) is shown in the ${\rm Re}[\rho]$--$ {\rm Im}[\rho]$ plane of Fig.\,\ref{fig:rho-B}, while Fig.\,\ref{fig:rho_a-B} shows $\rho'$ for the potential minimum for the superpotential (35).
In this case, NH of neutrino masses is only available.

\begin{figure}[h!]
	\begin{tabular}{ccc}
		\begin{minipage}{0.475\hsize}                                       
		\includegraphics[bb=0 0 400 280,width=\linewidth]{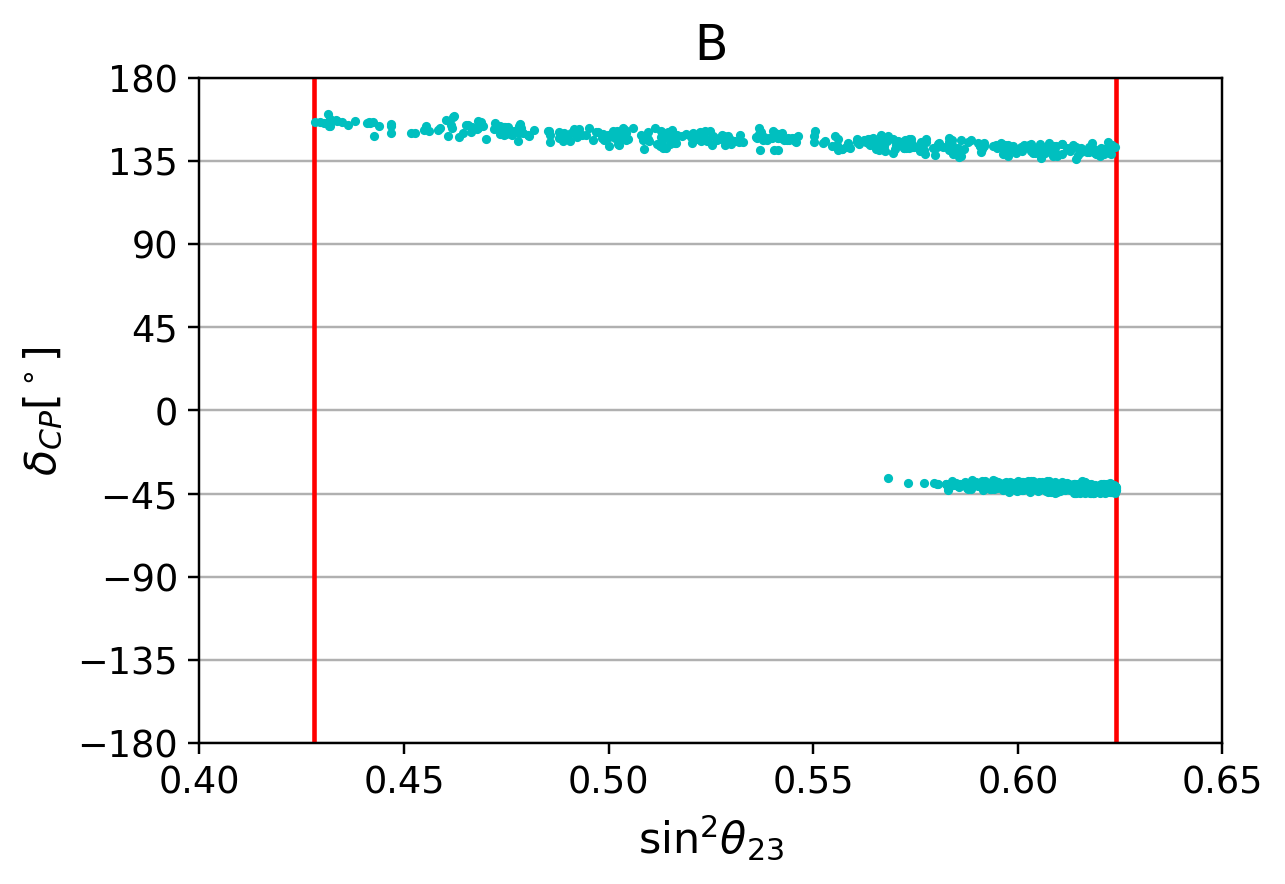}   
		\caption{The predicted region on $\sin^2\theta_{23}$                
		--$\delta_{CP}$ plane in B for NH.                                  
		Vertical red lines denote $3\sigma$ bound of observed data.}        
		\label{fig:23-delta_B}                                              
		\end{minipage}                                                      
		\phantom{=}                                                         
		\begin{minipage}{0.475\hsize}                                       
		\includegraphics[bb=0 0 400 280,width=\linewidth]{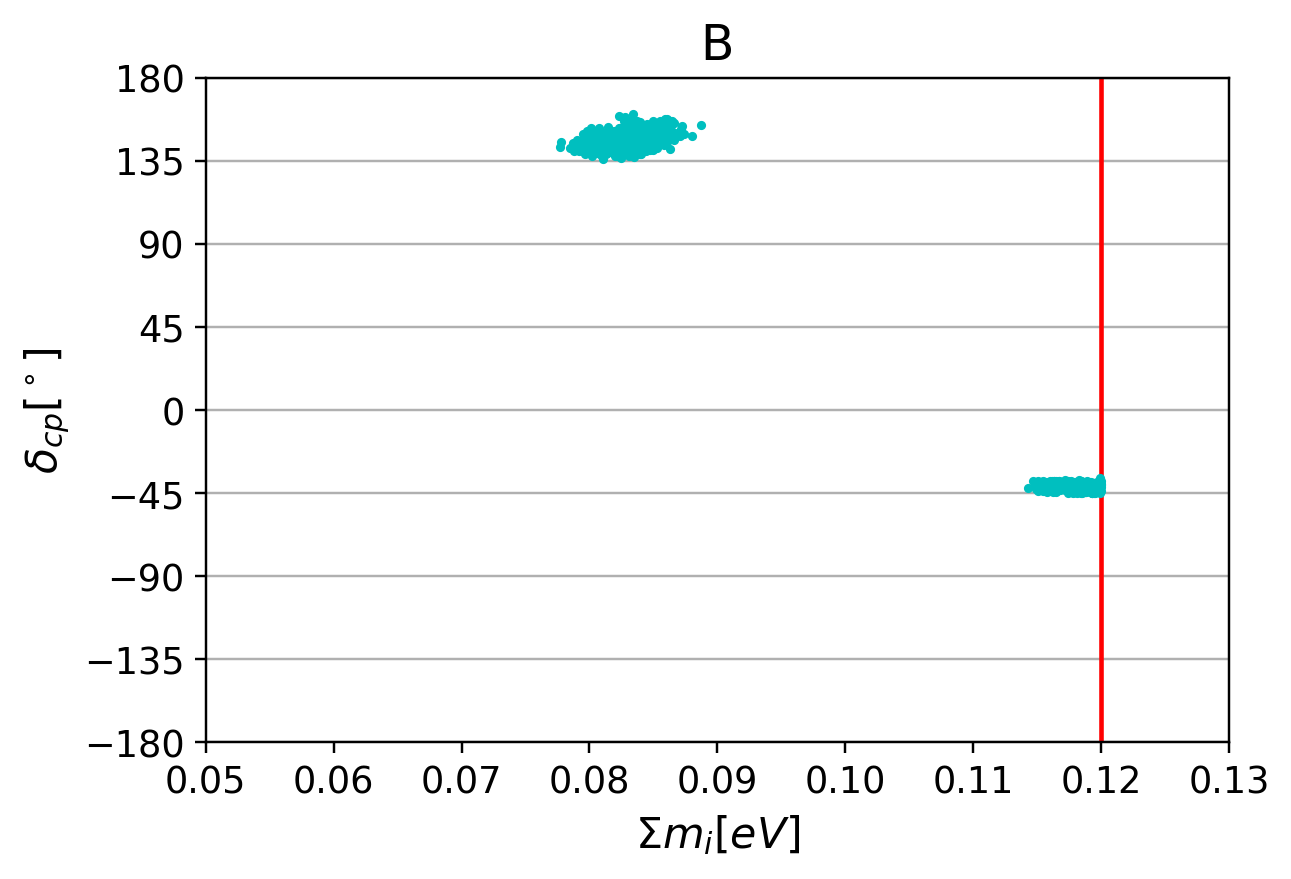} 
		\caption{The sum of neutrino masses $\sum m_i$                      
		dependence of $\delta_{CP}$ in B for NH.                            
		A vertical red line denote the cosmological bound. }                
		\label{fig:mass-delta_B}                                            
		\end{minipage}                                                      
	\end{tabular}
\end{figure}

We show the correlation between $\delta_{CP}$ and $\sin^2\theta_{23}$ in Fig.\,\ref{fig:23-delta_B}. 
The predicted ranges of $\delta_{CP}$ are $[140^\circ,160^\circ]$ for $\sin^2\theta_{23}\leq 0.57$
and $[135^\circ,140^\circ]$,  $[-45^\circ,-40^\circ]$ for $\sin^2\theta_{23}\geq 0.57$.
We present the predicted $\delta_{CP}$ versus the sum of neutrino masses $\sum m_i$ in Fig.\,\ref{fig:mass-delta_B}.
For $\sum m_i= [78, 88]$ [meV], $\delta_{CP}$ is predicted in
$[135^\circ,160^\circ]$.
Near the cosmological bound,  $\sum m_i= [115, 120]$ [meV],
$\delta_{CP}$ is $[-45^\circ,-40^\circ]$. 

\begin{figure}[h!]
	\begin{tabular}{ccc}
		\begin{minipage}{0.475\hsize}                                     
		\includegraphics[bb=0 0 400 280,width=\linewidth]{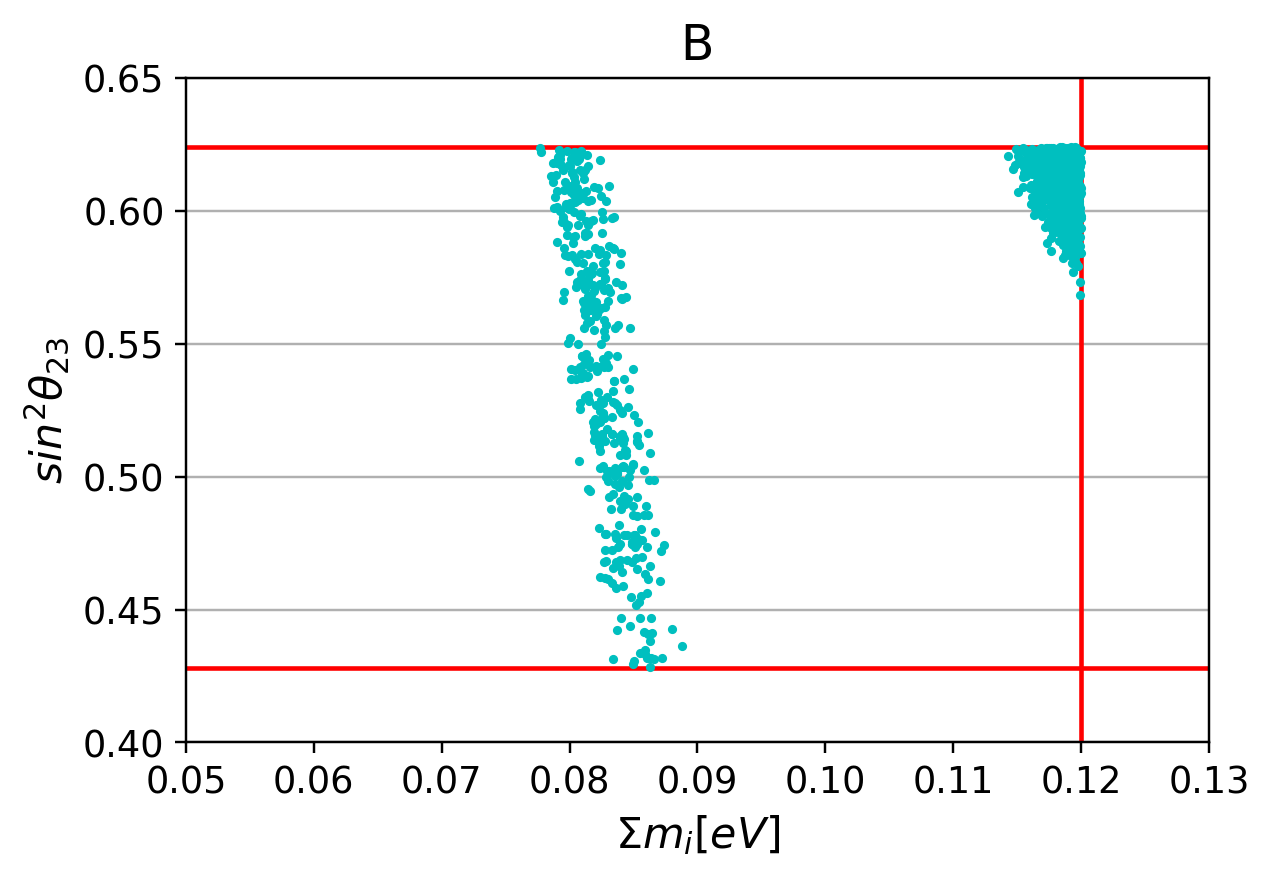}  
		\caption{The sum of neutrino masses $\sum m_i$                    
		dependence of $\sin^2\theta_{23}$ in B for NH. }                  
		\label{fig:mass-23_B}                                             
		\end{minipage}                                                    
		\phantom{=}                                                       
		\begin{minipage}{0.475\hsize}                                     
		\includegraphics[bb=0 0 400 280,width=\linewidth]{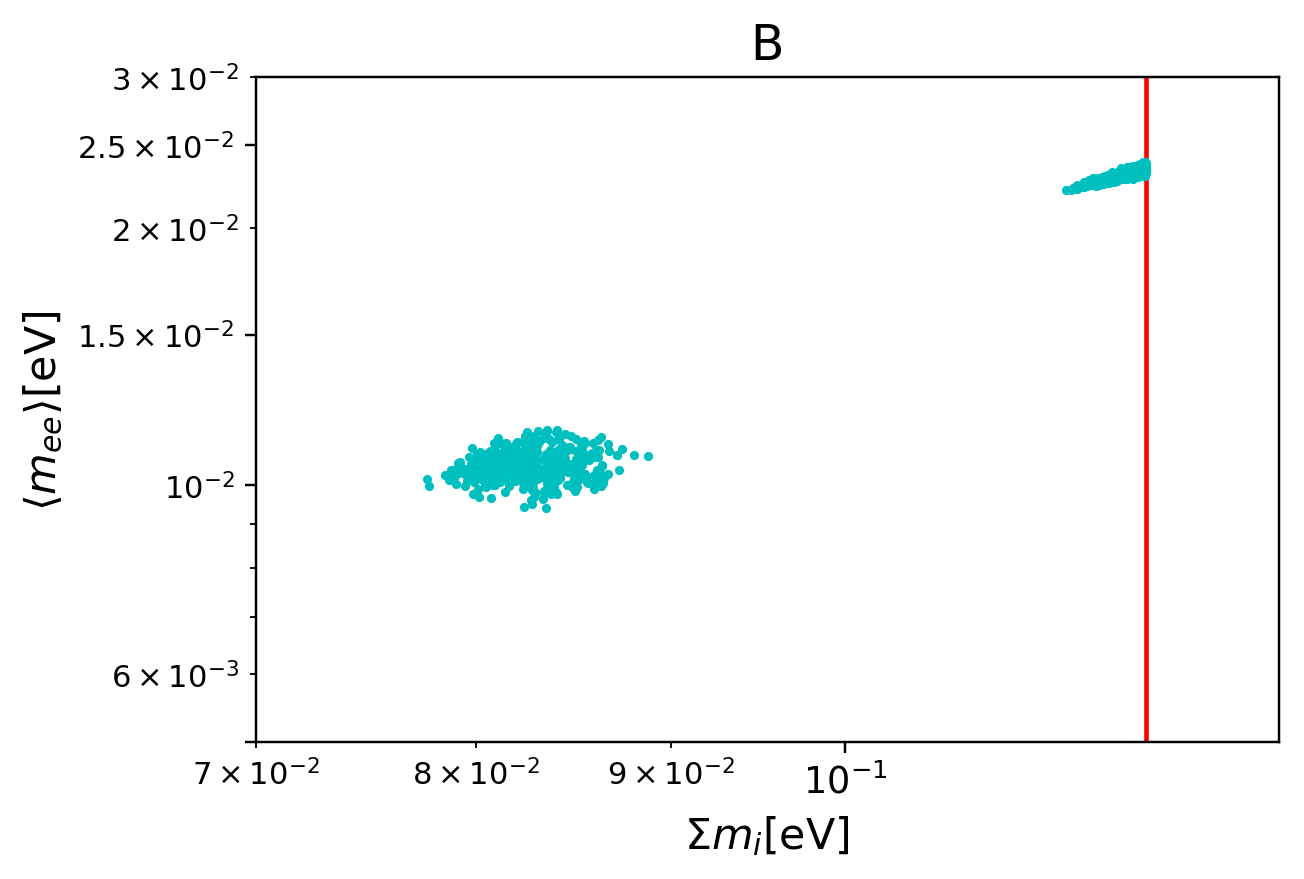} 
		\caption{The predicted  effective neutrino mass                   
		$\langle m_{ee}\rangle$ versus $\sum m_i$  in B for NH.}          
		\label{fig:mass-mee_B}                                            
		\end{minipage}                                                    
	\end{tabular}
\end{figure}

The neutrino mass dependence for $\sin^2\theta_{23}$ 
is almost same as the result of region $A$, as seen in Fig.\,\ref{fig:mass-23_B}.
The effective mass of the $0\nu\beta\beta$ decay $\langle m_{ee}\rangle$  is also same as the result of region $A$,
as seen in Fig.\,\ref{fig:mass-mee_B}. 
We summarize the prediction for $\langle m_{ee}\rangle$ and $\sum m_i$ in Table\,\ref{tb:ABC}.


\vskip 1 cm
\setcounter{table}{1}
\begin{table}[h!]
	\centering
	\begin{tabular}{|c||c|c|c|} \hline
		  & A                         & B                          & C              \\ \
		  & NH \qquad\qquad\quad\  IH & NH  \qquad\qquad\quad\  IH & NH  \qquad\ IH \\ \hline\hline
		\rule[14pt]{0pt}{0pt}
		$\langle m_{ee}\rangle$ [meV] & 9.4--12,\,  22--24 \quad\ \ \,
		$\times$ &  9.4--12,\,  22--24 \quad\ \ \, $\times$ & $\times$ \qquad 18--34\ \ \\
		$\sum m_i$ [meV] &77--88, 115--120 \ \ $\times$ & 77--88, 115--120 \ \ $\times$
		& $\times$ \quad\ \ 98--110 \\
		\hline
	\end{tabular}
	\caption{
		Predicted $\langle m_{ee}\rangle$ and $\sum m_i$ for cases A, B, and C. }
	\label{tb:ABC}
\end{table}

\subsection{Region of $C$}

Finally, we present numerical discussions in the region $C$ of  the modulus $\tau$.
The parameter $\rho$ which realizes the potential minimum for the superpotential (34) is shown in the ${\rm Re}[\rho]$--$ {\rm Im}[\rho]$ plane of Fig.\,\ref{fig:rho-C}, while Fig.\,\ref{fig:rho_a-C} shows $\rho'$ for the potential minimum for the superpotential (35).
In this case, IH is only available.

\begin{figure}[h!]
	\begin{tabular}{ccc}
		\begin{minipage}{0.475\hsize}                                       
		\includegraphics[bb=0 0 400 280,width=\linewidth]{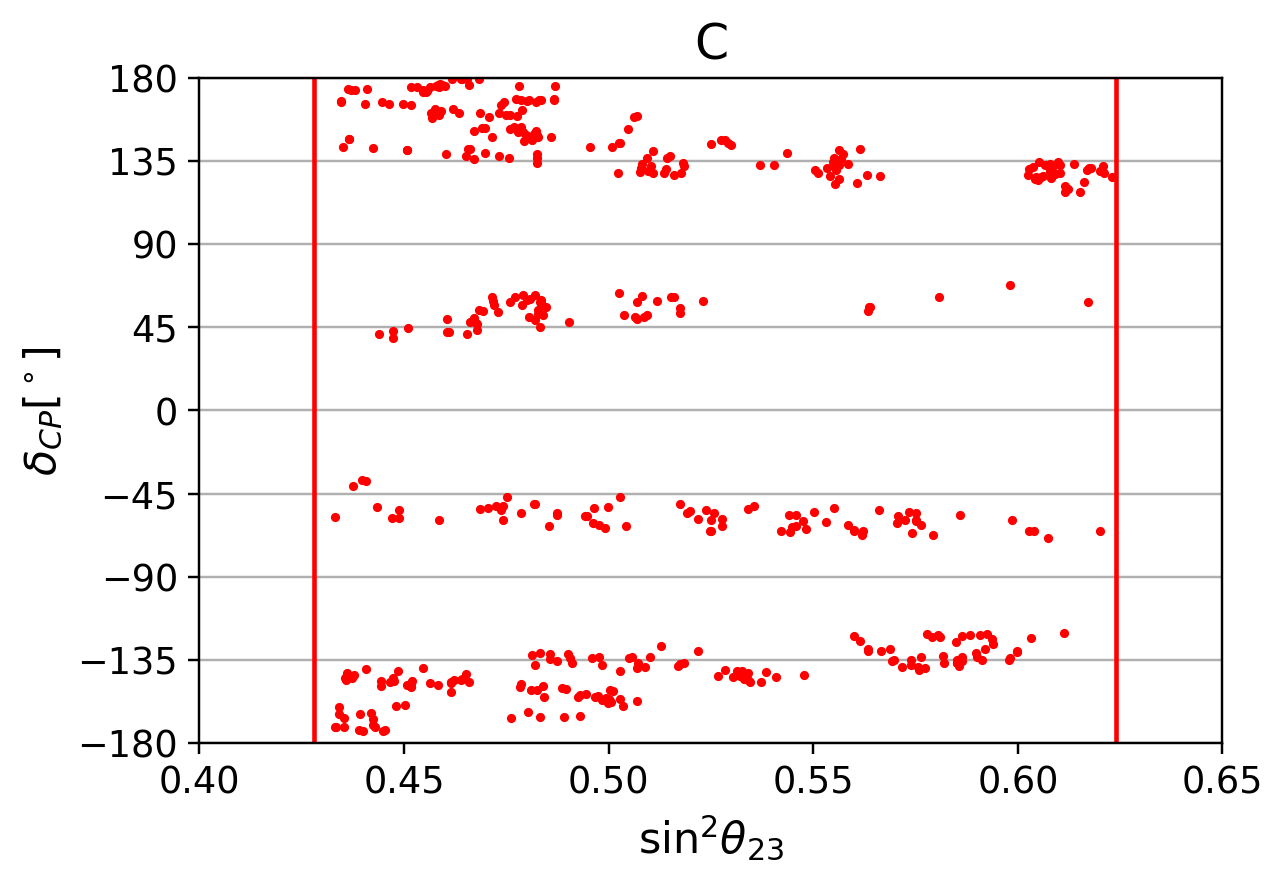}   
		\caption{The predicted region on $\sin^2\theta_{23}$                
		--$\delta_{CP}$ plane in C for IH.                                  
		Vertical red lines denote $3\sigma$ bound of observed data.}        
		\label{fig:23-delta_C}                                              
		\end{minipage}                                                      
		\phantom{=}                                                         
		\begin{minipage}{0.475\hsize}                                       
		\includegraphics[bb=0 0 400 280,width=\linewidth]{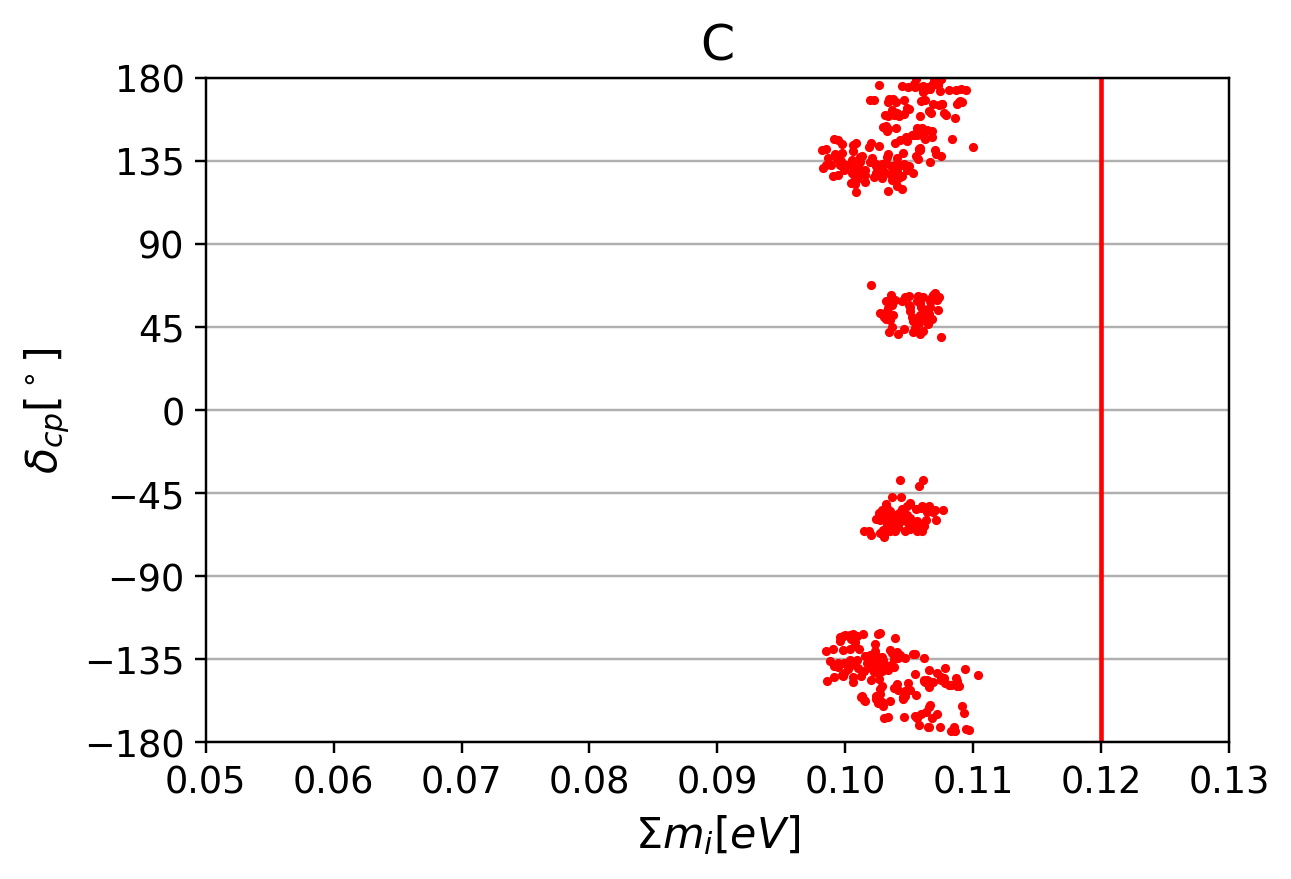} 
		\caption{The sum of neutrino masses $\sum m_i$                      
		dependence of $\delta_{CP}$ in C for IH.                            
		A vertical red line denote the cosmological bound. }                
		\label{fig:mass-delta_C}                                            
		\end{minipage}                                                      
	\end{tabular}
\end{figure}

We show the correlation between $\delta_{CP}$ and $\sin^2\theta_{23}$ in Fig.\,\ref{fig:23-delta_C}.
We can see the weak $\sin^2\theta_{23}$ dependence for $\delta_{CP}$.
The predicted $\delta_{CP}$ is in the range of
$\pm [40^\circ,60^\circ]$ and  $\pm [120^\circ,180^\circ]$.
We show the predicted $\delta_{CP}$ versus the sum of neutrino masses in Fig.\,\ref{fig:mass-delta_C}.
The sum of neutrino masses $\sum m_i$ is restricted in a narrow range, $\sum m_i= [98, 110]$ [meV].

\begin{figure}[h!]
	\begin{tabular}{ccc}
		\begin{minipage}{0.475\hsize}                                     
		\includegraphics[bb=0 0 400 280,width=\linewidth]{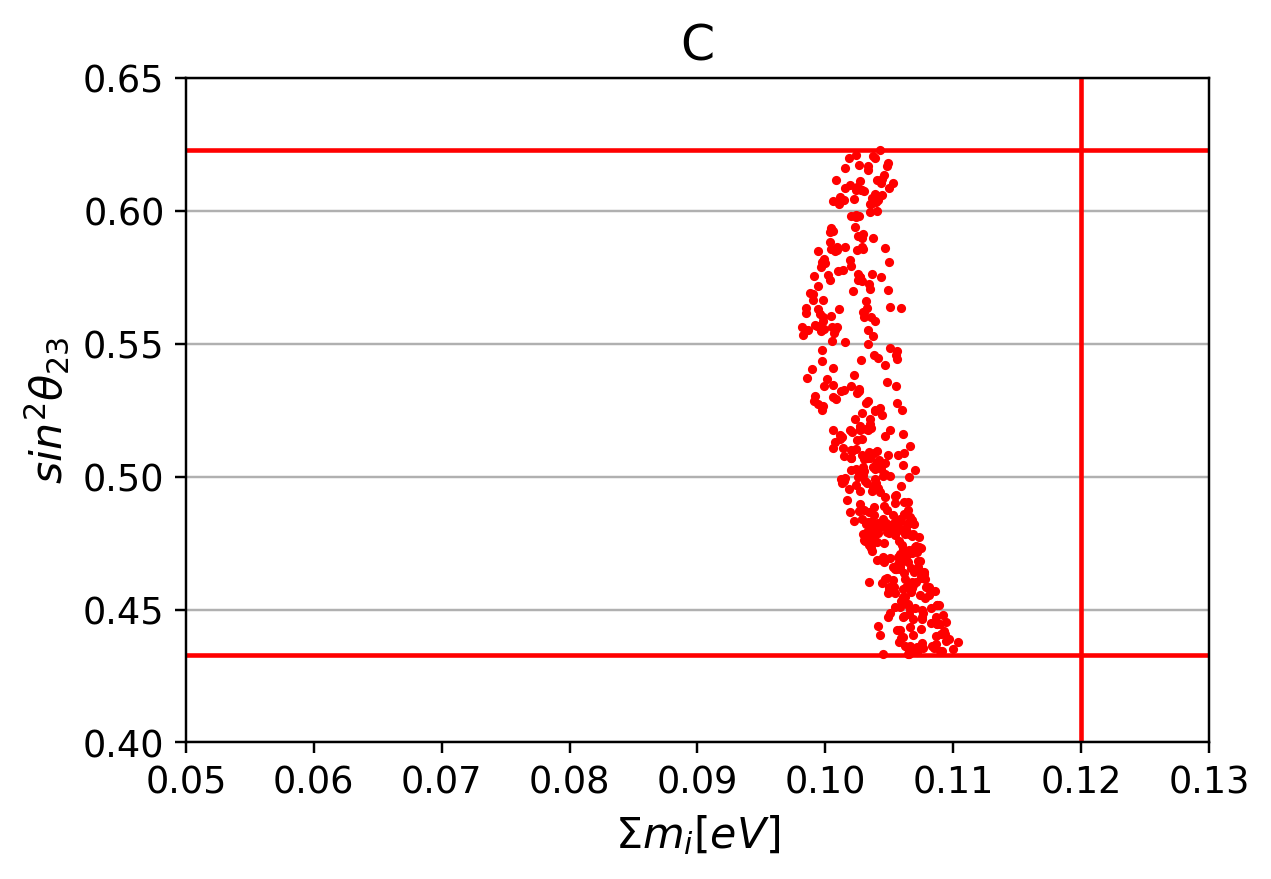}  
		\caption{The sum of neutrino masses $\sum m_i$                    
		dependence of $\sin^2\theta_{23}$ in C for IH.	}                  
		\label{fig:mass-23_C}                                             
		\end{minipage}                                                    
		\phantom{=}                                                       
		\begin{minipage}{0.475\hsize}                                     
		\includegraphics[bb=0 0 400 280,width=\linewidth]{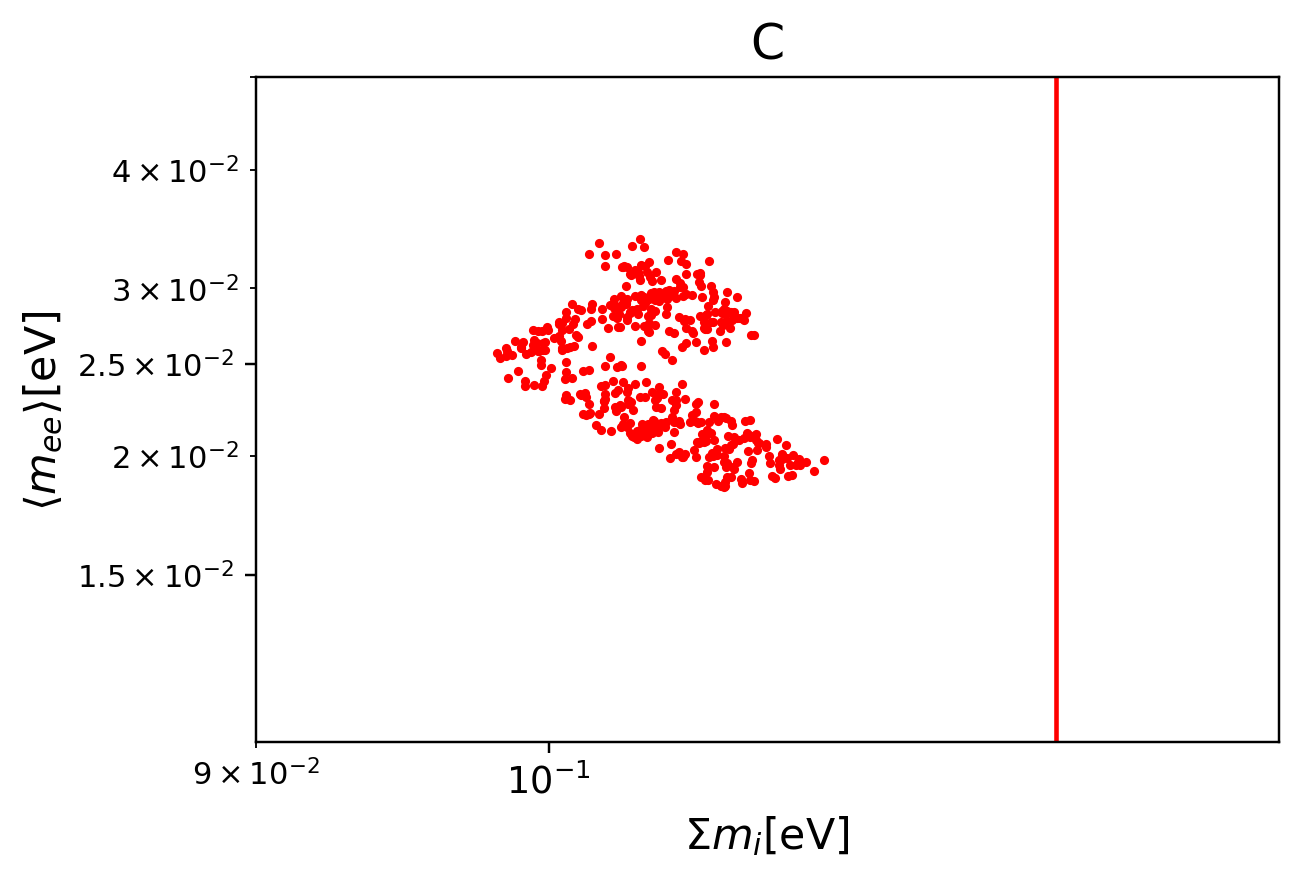} 
		\caption{The predicted  effective neutrino mass                   
		$\langle m_{ee}\rangle$ versus $\sum m_i$  in C for IH.}          
		\label{fig:mass-mee_C}                                            
		\end{minipage}                                                    
	\end{tabular}
\end{figure}

There is no clear neutrino mass dependence for $\sin^2\theta_{23}$ as seen in Fig.\,\ref{fig:mass-23_C}.
Below $\sum m_i= 102$ [meV], $\theta_{23}$ is predicted in the second octant while it is in the first octant in $\sum m_i\geq 107$ [meV].
The effective mass of the $0\nu\beta\beta$ decay $\langle m_{ee}\rangle$ is presented in Fig.\,\ref{fig:mass-mee_C}. The prediction is in  the range of  $18$--$34$ [meV].
We summarize the prediction for $\langle m_{ee}\rangle$ and $\sum m_i$ in Table \ref{tb:ABC}.


\section{Conclusion}

We have studied the modulus stabilization and its phenomenological aspects in the $A_4$ flavor model, where the $A_4$ flavor symmetry is originated from the $S_4$ modular symmetry.
We can stabilize the modulus by a superpotential with a single modular form, but its modulus value is not favorable 
in lepton masses and mixing angles in the $A_4$ flavor model.
If we assume two modular forms in the superpotential, we can stabilize the modulus at favorable values by using 
the parameter $\rho$ as well as $\rho'$.
Proper values of $\rho$ and $\rho'$ are of ${\cal O}(0.1-1)$.
Thus, contributions due to two modular forms are important in our model.
By choosing a proper value of $\rho$ as well as $\rho'$ in the superpotential, we can stabilize the value of $\tau$ in our scalar potential such that one can realize the lepton masses and its mixing angles.

We have presented the neutrino phenomenology in the three different regions of $\tau$ (A,B,C) where modulus stabilization is realized.
The CP violating phase of leptons, $\delta_{CP}$ is distinctly predicted in three regions of $\tau$.
It is also emphasized that IH of neutrino masses is reproduced in only C region.
The sum of neutrino masses is predicted in the restricted range for A, B and C respectively.
The cosmological observation of it will provide a crucial test of our model.
The effective mass of the $0\nu\beta\beta$ decay $\langle m_{ee}\rangle$ is also predicted. 
The future experiments can probe our model since our prediction includes $\langle m_{ee}\rangle = 22 \mbox{--} 24$ [meV] \cite{Gando:2019mxj}.   
Thus, our model realizes the modulus stabilization where the successful phenomenological results are obtained.

\vspace{1.5 cm}
\noindent
{\large\bf Acknowledgement}\\

This work is supported by  MEXT KAKENHI Grant Number JP19H04605 (TK), and 
JSPS Grants-in-Aid for Scientific Research  18J11233 (THT).
The work of YS is supported by JSPS KAKENHI Grant Number JP17K05418 and Fujyukai Foundation.

\appendix
\section*{Appendix}

\section{$S_4$ and $A_4$ representations}

The representations $S$ and $T$ of $\Gamma_4 \simeq S_4$ are given for the representations ${\bf 2}$ 
and ${\bf 3}'$ in section 2.
Here, we give other representations.
The generators $S$ and $T$ are represented by 
\begin{equation}
	\rho(S)=\frac13\left(
	\begin{array}{ccc}
		-1        & 2\omega^2 & 2 \omega  \\
		2\omega   & 2         & -\omega^2 \\
		2\omega^2 & -\omega   & 2         
	\end{array}\right), \qquad
	\rho(T)=
	\frac13\left(
	\begin{array}{ccc}
		-1        & 2\omega   & 2 \omega^2 \\
		2\omega   & 2\omega^2 & -1         \\
		2\omega^2 & -1        & 2\omega    
	\end{array}\right),
\end{equation}
on the $S_4$ ${\bf 3}$ representation, 
where $\omega=e^{i\frac{2}{3}\pi}$,
and 
\begin{equation}
	\rho(S) = \rho(T) =-1,
\end{equation}
for ${\bf 1}'$, while $\rho(S)=\rho(T)=1$ for ${\bf 1}$.

On the other hand, we take the generators of $A_4$ group as follows:
\begin{align}
	\begin{aligned}
	\rho(S)=\frac{1}{3}
	\begin{pmatrix}
	-1 & 2        & 2      \\
	2  & -1       & 2      \\
	2  & 2        & -1     
	\end{pmatrix},
	\end{aligned}
	\qquad 
	\begin{aligned}
	\rho(T)=
	\begin{pmatrix}
	1  & 0        & 0      \\
	0  & \omega^2 & 0      \\
	0  & 0        & \omega 
	\end{pmatrix}. 
	\end{aligned}
\end{align}
In this base, the multiplication rule of the $A_4$ triplet is
\begin{align}
	\begin{pmatrix}
	a_1\\
	a_2\\
	a_3
	\end{pmatrix}_{\bf 3}
	\otimes 
	\begin{pmatrix}
	b_1\\
	b_2\\
	b_3
	\end{pmatrix}_{\bf 3}
	                                             & =\left (a_1b_1+a_2b_3+a_3b_2\right )_{\bf 1} \oplus \left (a_3b_3+a_1b_2+a_2b_1\right )_{{\bf 1}'} \nonumber \\
	                                             & \oplus \left (a_2b_2+a_1b_3+a_3b_1\right )_{{\bf 1}''} \nonumber                                             \\
	                                             & \oplus \frac13                                                                                               
	\begin{pmatrix}
	2a_1b_1-a_2b_3-a_3b_2 \\
	2a_3b_3-a_1b_2-a_2b_1 \\
	2a_2b_2-a_1b_3-a_3b_1
	\end{pmatrix}_{{\bf 3}}
	\oplus \frac12
	\begin{pmatrix}
	a_2b_3-a_3b_2 \\
	a_1b_2-a_2b_1 \\
	a_3b_1-a_1b_3
	\end{pmatrix}_{{\bf 3}\  } \ , \nonumber \\
	\nonumber \\
	{\bf 1} \otimes {\bf 1} = {\bf 1} \ , \qquad &                                                                                                              
	{\bf 1'} \otimes {\bf 1'} = {\bf 1''} \ , \qquad
	{\bf 1''} \otimes {\bf 1''} = {\bf 1'} \ , \qquad
	{\bf 1'} \otimes {\bf 1''} = {\bf 1} \  .
\end{align}

More details are shown in the review~\cite{Ishimori:2010au,Ishimori:2012zz}.

\section{Input data}
We input charged lepton masses in order to constrain the model parameters.
We take Yukawa couplings of charged leptons 
at the GUT scale $2\times 10^{16}$ GeV,  where $\tan\beta=2.5$ is taken
\cite{Criado:2018thu,Antusch:2005gp,Antusch:2013jca,Bjorkeroth:2015ora}:
\begin{eqnarray}
	y_e=(1.97\pm 0.02) \times 10^{-6}, \quad 
	y_\mu=(4.16\pm 0.05) \times 10^{-4}, \quad 
	y_\tau=(7.07\pm 0.07) \times 10^{-3},
\end{eqnarray}
where lepton masses are  given by $m_\ell=\sqrt{2} y_\ell v_H$ with $v_H=174$ GeV.
We also use the following lepton mixing angles and neutrino mass parameters in Table 2 given by NuFIT 4.0 \cite{NuFIT}.
The RGE effects of mixing angles and the mass ratio $\Delta m_{\rm sol}^2/\Delta m_{\rm atm}^2$
are negligibly small in the case of $\tan\beta=2.5$ for both NH and IH as seen in Appendix E of Ref.\,\cite{Criado:2018thu}.
\begin{table}[h]
	\begin{center}
		\begin{tabular}{|c|c|c|}
			\hline 
			\rule[14pt]{0pt}{0pt}
			\  observable \&$3\,\sigma$ range for NH & $3\,\sigma$ range for IH &                      \\
			\hline 
			\rule[14pt]{0pt}{0pt}
			$\Delta m_{\rm atm}^2$& \ \   \ \ $(2.431$--$ 2.622) \times 10^{-3}{\rm eV}^2$ \ \ \ \
			&\ \ $- (2.413$--$2.606) \times 10^{-3}{\rm eV}^2$ \ \  \\
			\hline 
			\rule[14pt]{0pt}{0pt}
			$\Delta m_{\rm sol }^2$& $(6.79$--$ 8.01) \times 10^{-5}{\rm eV}^2$
			& $(6.79$--$ 8.01)  \times 10^{-5}{\rm eV}^2$ \\
			\hline 
			\rule[14pt]{0pt}{0pt}
			$\sin^2\theta_{23}$                      & $0.428$--$ 0.624$        & $0.433$--$ 0.623$    \\
			\hline 
			\rule[14pt]{0pt}{0pt}
			$\sin^2\theta_{12}$                      & $0.275$--$ 0.350$        & $0.275$--$ 0.350$    \\
			\hline 
			\rule[14pt]{0pt}{0pt}
			$\sin^2\theta_{13}$                      & $0.02044$--$ 0.02437$    & $0.02067$--$0.02461$ \\
			\hline 
		\end{tabular}
		\caption{The $3\,\sigma$ ranges of neutrino  parameters from NuFIT 4.0
			for NH and IH 
			\cite{NuFIT}. 
		}
		\label{DataNufit}
	\end{center}
\end{table}


\section{Modulus potential}

We give a scenario on a plausible mechanism to induce 
the modulus superpotential.
We assume a hidden sector, e.g. supersymmetric QCD which has the $SU(N)$ gauge symmetry and  $N_f$ flavors of  chiral matter fields, $Q_i$ and $\bar  Q^j$, with fundamental and anti-fundamental representations.
When $N=N_f$, the mesons $M=Q_i\bar Q^j$ and the baryon $B=\varepsilon^{i_1 \cdots i_N}Q_{i_1} \cdots Q_{i_N}$ 
as well as anti-baryon condensate \cite{Intriligator:1995au}.
If the superpotential at the tree level has the mass term, $W=m^i_iQ_i\bar Q^j$, the above condensation 
leads to the term $W=m \langle M \rangle$.
Suppose that the $M$ has the modular weight $-k-1$.
The mass parameter must be a modular form of the weight $k$ since the modular invariance requires that $m$ has the modular weight $k$.
The following superpotential may be induced 
\begin{equation}
	W = cY^{(k)}(\tau) \langle M \rangle .
\end{equation}
If there is another hidden sector to condensate, we may realize the superpotential
\begin{equation}
	W = cY^{(k)}(\tau) \langle M \rangle + c'Y^{(k')}(\tau) \langle M' \rangle ,
\end{equation}
where we assume that another meson fields $M'$ has the modular weight $-k'-1$.
Furthermore, the condensation of the baryon $B$ may induce another term with a different modular weight.


\begin{thebibliography}{99}
					
					
					
	\bibitem{Abe:2018wpn}
	K.~Abe {\it et al.} [T2K Collaboration],
	Phys.\ Rev.\ Lett.\  {\bf 121} (2018) no.17,  171802
	[arXiv:1807.07891 [hep-ex]].
					
	\bibitem{NOvA:2018gge}
	M.~A.~Acero {\it et al.} [NOvA Collaboration],
	Phys.\ Rev.\ D {\bf 98} (2018) 032012
	[arXiv:1806.00096 [hep-ex]].
					
					
					
	\bibitem{Altarelli:2010gt}
	G.~Altarelli and F.~Feruglio,
	Rev.\ Mod.\ Phys.\  {\bf 82} (2010) 2701
	[arXiv:1002.0211 [hep-ph]].
					
					
					
	\bibitem{Ishimori:2010au}
	H.~Ishimori, T.~Kobayashi, H.~Ohki, Y.~Shimizu, H.~Okada and M.~Tanimoto,
	Prog.\ Theor.\ Phys.\ Suppl.\  {\bf 183} (2010) 1
	[arXiv:1003.3552 [hep-th]].
					
					
					
	\bibitem{Ishimori:2012zz}
	H.~Ishimori, T.~Kobayashi, H.~Ohki, H.~Okada, Y.~Shimizu and M.~Tanimoto,
	Lect.\ Notes Phys.\  {\bf 858} (2012) 1, Springer.
					
	\bibitem{Hernandez:2012ra}
	D.~Hernandez and A.~Y.~Smirnov,
	Phys.\ Rev.\ D {\bf 86} (2012) 053014
	[arXiv:1204.0445 [hep-ph]].
					
	\bibitem{King:2013eh}
	S.~F.~King and C.~Luhn,
	Rept.\ Prog.\ Phys.\  {\bf 76} (2013) 056201
	[arXiv:1301.1340 [hep-ph]].
					
	\bibitem{King:2014nza} 
	S.~F.~King, A.~Merle, S.~Morisi, Y.~Shimizu and M.~Tanimoto,
	New J.\ Phys.\  {\bf 16}, 045018 (2014)
	[arXiv:1402.4271 [hep-ph]].
					
					
	\bibitem{Tanimoto:2015nfa}
	M.~Tanimoto,
	AIP Conf.\ Proc.\  {\bf 1666} (2015) 120002.
					
	\bibitem{King:2017guk}
	S.~F.~King,
	Prog.\ Part.\ Nucl.\ Phys.\  {\bf 94} (2017) 217
	[arXiv:1701.04413 [hep-ph]].
					
	\bibitem{Petcov:2017ggy}
	S.~T.~Petcov,
	Eur.\ Phys.\ J.\ C {\bf 78} (2018) no.9,  709
	[arXiv:1711.10806 [hep-ph]].
					
					  
	\bibitem{Ma:2001dn}
	E.~Ma and G.~Rajasekaran,
	Phys.\ Rev.\  D {\bf 64}, 113012 (2001)
	[arXiv:hep-ph/0106291].
					
					
					
	\bibitem{Babu:2002dz}
	K.~S.~Babu, E.~Ma and J.~W.~F.~Valle,
	Phys.\ Lett.\  B {\bf 552}, 207 (2003)
	[arXiv:hep-ph/0206292].
					
	\bibitem{Altarelli:2005yp}
	G.~Altarelli and F.~Feruglio,
	Nucl.\ Phys.\ B {\bf 720} (2005) 64
	[hep-ph/0504165].
					
					
	\bibitem{Altarelli:2005yx}
	G.~Altarelli and F.~Feruglio,
	Nucl.\ Phys.\ B {\bf 741} (2006) 215
	[hep-ph/0512103].
					
	\bibitem{Shimizu:2011xg}
	Y.~Shimizu, M.~Tanimoto and A.~Watanabe,
	Prog.\ Theor.\ Phys.\  {\bf 126} (2011) 81
	[arXiv:1105.2929 [hep-ph]].
					
	\bibitem{Petcov:2018snn}
	S.~T.~Petcov and A.~V.~Titov,
	Phys.\ Rev.\ D {\bf 97} (2018) no.11,  115045
	[arXiv:1804.00182 [hep-ph]].
					
	\bibitem{Kang:2018txu} 
	S.~K.~Kang, Y.~Shimizu, K.~Takagi, S.~Takahashi and M.~Tanimoto,
	PTEP {\bf 2018}, no. 8, 083B01 (2018)
	[arXiv:1804.10468 [hep-ph]].
					
					
	\bibitem{Kobayashi:2006wq}
	T.~Kobayashi, H.~P.~Nilles, F.~Ploger, S.~Raby and M.~Ratz,
	Nucl.\ Phys.\ B {\bf 768}, 135 (2007)
	[hep-ph/0611020].
					
					
	\bibitem{Kobayashi:2004ya}
	T.~Kobayashi, S.~Raby and R.~J.~Zhang,
	Nucl.\ Phys.\ B {\bf 704}, 3 (2005)
	[hep-ph/0409098].
					
					
	\bibitem{Ko:2007dz}
	P.~Ko, T.~Kobayashi, J.~h.~Park and S.~Raby,
	Phys.\ Rev.\ D {\bf 76}, 035005 (2007)
	Erratum: [Phys.\ Rev.\ D {\bf 76}, 059901 (2007)]
	[arXiv:0704.2807 [hep-ph]].
					
	\bibitem{Beye:2014nxa} 
	F.~Beye, T.~Kobayashi and S.~Kuwakino,
	Phys.\ Lett.\ B {\bf 736}, 433 (2014)
	[arXiv:1406.4660 [hep-th]].
					
					
	\bibitem{Abe:2009vi}
	H.~Abe, K.~S.~Choi, T.~Kobayashi and H.~Ohki,
	Nucl.\ Phys.\ B {\bf 820}, 317 (2009)
	[arXiv:0904.2631 [hep-ph]].
					
					
					
					
					
	\bibitem{deAdelhartToorop:2011re} 
	R.~de Adelhart Toorop, F.~Feruglio and C.~Hagedorn,
	Nucl.\ Phys.\ B {\bf 858}, 437 (2012)
	[arXiv:1112.1340 [hep-ph]].
					
					
	\bibitem{Lauer:1989ax} 
	J.~Lauer, J.~Mas and H.~P.~Nilles,
	Phys.\ Lett.\ B {\bf 226}, 251 (1989).
	%
	Nucl.\ Phys.\ B {\bf 351}, 353 (1991).
					  
	\bibitem{Lerche:1989cs} 
	W.~Lerche, D.~Lust and N.~P.~Warner,
	Phys.\ Lett.\ B {\bf 231}, 417 (1989).
					
	\bibitem{Ferrara:1989qb} 
	S.~Ferrara, .D.~Lust and S.~Theisen,
	Phys.\ Lett.\ B {\bf 233}, 147 (1989).
					  
	\bibitem{Kobayashi:2017dyu} 
	T.~Kobayashi and S.~Nagamoto,
	Phys.\ Rev.\ D {\bf 96}, no. 9, 096011 (2017)
	[arXiv:1709.09784 [hep-th]].
					  
	\bibitem{Kobayashi:2018rad} 
	T.~Kobayashi, S.~Nagamoto, S.~Takada, S.~Tamba and T.~H.~Tatsuishi,
	Phys.\ Rev.\ D {\bf 97}, no. 11, 116002 (2018)
	[arXiv:1804.06644 [hep-th]].
					 
	\bibitem{Baur:2019kwi} 
	A.~Baur, H.~P.~Nilles, A.~Trautner and P.~K.~S.~Vaudrevange,
	Phys.\ Lett.\ B {\bf 795}, 7 (2019)
	[arXiv:1901.03251 [hep-th]]; 
	%
	arXiv:1908.00805 [hep-th].
					  
					  
					
					
					
	\bibitem{Feruglio:2017spp}
	F.~Feruglio,
	arXiv:1706.08749 [hep-ph].
					
	\bibitem{Criado:2018thu}
	J.~C.~Criado and F.~Feruglio,
	SciPost Phys.\  {\bf 5} (2018) no.5,  042
	[arXiv:1807.01125 [hep-ph]].
					
					
					
	\bibitem{Kobayashi:2018scp}
	T.~Kobayashi, N.~Omoto, Y.~Shimizu, K.~Takagi, M.~Tanimoto and T.~H.~Tatsuishi,
	JHEP {\bf 1811} (2018) 196
	[arXiv:1808.03012 [hep-ph]].
					
	\bibitem{Kobayashi:2018vbk}
	T.~Kobayashi, K.~Tanaka and T.~H.~Tatsuishi,
	Phys.\ Rev.\ D {\bf 98} (2018) no.1,  016004
	[arXiv:1803.10391 [hep-ph]].
					
					
	\bibitem{Penedo:2018nmg}
	J.~T.~Penedo and S.~T.~Petcov,
	Nucl.\ Phys.\ B {\bf 939} (2019) 292
	[arXiv:1806.11040 [hep-ph]].
					
	\bibitem{Novichkov:2018nkm}
	P.~P.~Novichkov, J.~T.~Penedo, S.~T.~Petcov and A.~V.~Titov,
	JHEP {\bf 1904} (2019) 174
	[arXiv:1812.02158 [hep-ph]].
					
	\bibitem{Kobayashi:2018bff}
	T.~Kobayashi and S.~Tamba,
	Phys.\ Rev.\ D {\bf 99} (2019) no.4,  046001
	[arXiv:1811.11384 [hep-th]].
					
					
	\bibitem{Liu:2019khw}
	X.~G.~Liu and G.~J.~Ding,
	JHEP {\bf 1908} (2019) 134
	[arXiv:1907.01488 [hep-ph]].
					
					
	\bibitem{Novichkov:2018ovf}
	P.~P.~Novichkov, J.~T.~Penedo, S.~T.~Petcov and A.~V.~Titov,
	JHEP {\bf 1904} (2019) 005
	[arXiv:1811.04933 [hep-ph]].
					
					
					
	\bibitem{deAnda:2018ecu}
	F.~J.~de Anda, S.~F.~King and E.~Perdomo,
	arXiv:1812.05620 [hep-ph].
					
	\bibitem{Okada:2018yrn}
	H.~Okada and M.~Tanimoto,
	Phys.\ Lett.\ B {\bf 791} (2019) 54
	[arXiv:1812.09677 [hep-ph]].
					
					
	\bibitem{Kobayashi:2018wkl} 
	T.~Kobayashi, Y.~Shimizu, K.~Takagi, M.~Tanimoto, T.~H.~Tatsuishi and H.~Uchida,
	Phys.\ Lett.\ B {\bf 794}, 114 (2019)
	[arXiv:1812.11072 [hep-ph]].
					
					
	\bibitem{Novichkov:2018yse}
	P.~P.~Novichkov, S.~T.~Petcov and M.~Tanimoto,
	Phys.\ Lett.\ B {\bf 793} (2019) 247
	[arXiv:1812.11289 [hep-ph]].
					
					
					
	\bibitem{Ding:2019xna}
	G.~J.~Ding, S.~F.~King and X.~G.~Liu,
	arXiv:1903.12588 [hep-ph].
					
					
					
	\bibitem{Nomura:2019jxj} 
	T.~Nomura and H.~Okada,
	Phys.\ Lett.\ B {\bf 797}, 134799 (2019)
	[arXiv:1904.03937 [hep-ph]].
					
					
	\bibitem{Novichkov:2019sqv} 
	P.~P.~Novichkov, J.~T.~Penedo, S.~T.~Petcov and A.~V.~Titov,
	JHEP {\bf 1907}, 165 (2019)
	[arXiv:1905.11970 [hep-ph]].
					
					
	\bibitem{Okada:2019uoy} 
	H.~Okada and M.~Tanimoto,
	arXiv:1905.13421 [hep-ph].
					
					
	\bibitem{deMedeirosVarzielas:2019cyj}
	I.~de Medeiros Varzielas, S.~F.~King and Y.~L.~Zhou,
	arXiv:1906.02208 [hep-ph].
					
					
	\bibitem{Nomura:2019yft}
	T.~Nomura and H.~Okada,
	arXiv:1906.03927 [hep-ph].
					
					
					
	\bibitem{Kobayashi:2019rzp} 
	T.~Kobayashi, Y.~Shimizu, K.~Takagi, M.~Tanimoto and T.~H.~Tatsuishi,
	arXiv:1906.10341 [hep-ph].
					
					
					
					
	\bibitem{Okada:2019xqk}
	H.~Okada and Y.~Orikasa,
	arXiv:1907.04716 [hep-ph].
					
					
	\bibitem{Kobayashi:2019mna} 
	T.~Kobayashi, Y.~Shimizu, K.~Takagi, M.~Tanimoto and T.~H.~Tatsuishi,
	arXiv:1907.09141 [hep-ph].
					
					
	\bibitem{Ding:2019zxk} 
	G.~J.~Ding, S.~F.~King and X.~G.~Liu,
	arXiv:1907.11714 [hep-ph].
					  
	\bibitem{Okada:2019mjf} 
	H.~Okada and Y.~Orikasa,
	arXiv:1907.13520 [hep-ph].
					  
					
	\bibitem{King:2019vhv} 
	S.~F.~King and Y.~L.~Zhou,
	arXiv:1908.02770 [hep-ph].
					  
	\bibitem{Nomura:2019lnr}
	T.~Nomura, H.~Okada and O.~Popov,
	arXiv:1908.07457 [hep-ph].
					  
	\bibitem{Okada:2019lzv}
	H.~Okada and Y.~Orikasa,
	arXiv:1908.08409 [hep-ph].
					  
	\bibitem{Criado:2019tzk}
	J.~C.~Criado, F.~Feruglio and S.~J.~D.~King,
	arXiv:1908.11867 [hep-ph].
					
					
	\bibitem{Ferrara:1989bc} 
	S.~Ferrara, D.~Lust, A.~D.~Shapere and S.~Theisen,
	Phys.\ Lett.\ B {\bf 225}, 363 (1989).
					  
	\bibitem{Derendinger:1991hq} 
	J.~P.~Derendinger, S.~Ferrara, C.~Kounnas and F.~Zwirner,
	Nucl.\ Phys.\ B {\bf 372}, 145 (1992).
					  
	\bibitem{Ibanez:1992hc} 
	L.~E.~Ibanez and D.~Lust,
	Nucl.\ Phys.\ B {\bf 382}, 305 (1992)
	[hep-th/9202046].
					  
	\bibitem{Kobayashi:2016ovu} 
	T.~Kobayashi, S.~Nagamoto and S.~Uemura,
	PTEP {\bf 2017}, no. 2, 023B02 (2017)
	[arXiv:1608.06129 [hep-th]].
					  
	\bibitem{Ferrara:1990ei} 
	S.~Ferrara, N.~Magnoli, T.~R.~Taylor and G.~Veneziano,
	Phys.\ Lett.\ B {\bf 245}, 409 (1990).
					  
	\bibitem{Cvetic:1991qm} 
	M.~Cvetic, A.~Font, L.~E.~Ibanez, D.~Lust and F.~Quevedo,
	Nucl.\ Phys.\ B {\bf 361}, 194 (1991).
					  
					
	\bibitem{Kobayashi:2016mzg} 
	T.~Kobayashi, D.~Nitta and Y.~Urakawa,
	JCAP {\bf 1608}, no. 08, 014 (2016)
	[arXiv:1604.02995 [hep-th]].
					
					
					
					
					
	\bibitem{Araki:2008ek}
	T.~Araki, T.~Kobayashi, J.~Kubo, S.~Ramos-Sanchez, M.~Ratz and P.~K.~S.~Vaudrevange,
	Nucl.\ Phys.\ B {\bf 805} (2008) 124
	[arXiv:0805.0207 [hep-th]].
					
	\bibitem{Kariyazono:2019ehj} 
	Y.~Kariyazono, T.~Kobayashi, S.~Takada, S.~Tamba and H.~Uchida,
	Phys.\ Rev.\ D {\bf 100}, no. 4, 045014 (2019)
	[arXiv:1904.07546 [hep-th]].
					
	\bibitem{Krauss:1988zc} 
	L.~M.~Krauss and F.~Wilczek,
	Phys.\ Rev.\ Lett.\  {\bf 62}, 1221 (1989).
					
					
	\bibitem{Ibanez:1991hv} 
	L.~E.~Ibanez and G.~G.~Ross,
	Phys.\ Lett.\ B {\bf 260}, 291 (1991).
					
	\bibitem{Banks:1991xj} 
	T.~Banks and M.~Dine,
	Phys.\ Rev.\ D {\bf 45}, 1424 (1992)
	[hep-th/9109045].
					
					
					
	\bibitem{NuFIT}
	NuFIT 4.0 (2018), www.nu-fit.org/;\\
	I.~Esteban, M.~C.~Gonzalez-Garcia, M.~Maltoni, I.~Martinez-Soler and T.~Schwetz,
	JHEP {\bf 1701}, 087 (2017)
	[arXiv:1611.01514 [hep-ph]].
	\bibitem{Tanabashi:2018oca}
	M.~Tanabashi {\it et al.} [Particle Data Group],
	Phys.\ Rev.\ D {\bf 98} (2018) no.3,  030001.
	\bibitem{Vagnozzi:2017ovm}
	S.~Vagnozzi, E.~Giusarma, O.~Mena, K.~Freese, M.~Gerbino, S.~Ho and M.~Lattanzi,
	Phys.\ Rev.\ D {\bf 96} (2017) no.12,  123503
	[arXiv:1701.08172 [astro-ph.CO]].
					
	\bibitem{Aghanim:2018eyx}
	N.~Aghanim {\it et al.} [Planck Collaboration],
	arXiv:1807.06209 [astro-ph.CO].
	\bibitem{Gando:2019mxj}
	Y.~Gando [KamLAND-Zen Collaboration],
	arXiv:1904.06655 [physics.ins-det].
	\bibitem{Antusch:2005gp}
	S.~Antusch, J.~Kersten, M.~Lindner, M.~Ratz and M.~A.~Schmidt,
	JHEP {\bf 0503} (2005) 024
	[hep-ph/0501272].
					
	\bibitem{Antusch:2013jca}
	S.~Antusch and V.~Maurer,
	JHEP {\bf 1311} (2013) 115
	[arXiv:1306.6879 [hep-ph]].
					
	\bibitem{Bjorkeroth:2015ora}
	F.~Bj\"orkeroth, F.~J.~de Anda, I.~de Medeiros Varzielas and S.~F.~King,
	JHEP {\bf 1506} (2015) 141
	[arXiv:1503.03306 [hep-ph]].
					
					
	\bibitem{Intriligator:1995au} 
	K.~A.~Intriligator and N.~Seiberg,
	Nucl.\ Phys.\ Proc.\ Suppl.\  {\bf 45BC}, 1 (1996)
	[Subnucl.\ Ser.\  {\bf 34}, 237 (1997)]
	[hep-th/9509066].
					
					
					
\end{thebibliography}
\end{document}